\RequirePackage{tikz}

\documentclass[aps,prx,reprint,superscriptaddress,nofootinbib]{revtex4-1}

\usepackage{tikz}
\usepackage{amsmath}
\usepackage{amssymb}
\usepackage{bm}
\usepackage{verbatim}
\usepackage{graphicx}
\usepackage{appendix}

\usetikzlibrary{%
	calc,%
	decorations.pathmorphing,%
	fadings,%
	shadings%
}

\usepackage{psfrag}
\usepackage{epsfig}
\usepackage{mathrsfs}

\newcommand{\eq}[1]{\begin{align}#1\end{align}}

\newcommand{\ffrac}[2]{\mbox{$\frac{#1}{#2}$}}
\def\half{\mbox{$\frac{1}{2}$}}

\def\tr{\mbox{tr}}

\newcommand{\p}{\partial}

\def\rb{\bm{r}}
\def\qb{\bm{q}}

\def\nb{\bm{n}}

\def\delb{\bm{\hat{\delta}}}
\def\Gb{\bm{G}}

\def\Rb{\bm{R}}

\def\curlyM{\bm{\mathcal{M}}}
\def\curlyP{\mathcal{P}}

\def\curlyT{\bm{\mathcal{T}}}

\def\curlyO{\mathcal{O}}

\newcommand{\dz}{\delta z}
\newcommand{\OO}{\mathcal{O}}
\newcommand\Gbo{\overline{\Gb}}

\def\kpa{k^\parallel}
\def\kpe{k^\perp}

\begin{document}

\title{Effects of coordination and pressure on sound attenuation, boson peak and elasticity in amorphous solids}

\author{Eric DeGiuli}
\affiliation{Center for Soft Matter Research, New York University, 4 Washington Place, New York, NY, 10003, USA}
\author{Adrien Laversanne-Finot}
\affiliation{Center for Soft Matter Research, New York University, 4 Washington Place, New York, NY, 10003, USA}
\author{Gustavo D\"uring}
\affiliation{Center for Soft Matter Research, New York University, 4 Washington Place, New York, NY, 10003, USA}
\affiliation{Facultad de F\'isica, Pontificia Universidad Cat\'olica de Chile, Casilla 306, Santiago 22, Chile}
\author{Edan Lerner}
\affiliation{Center for Soft Matter Research, New York University, 4 Washington Place, New York, NY, 10003, USA}
\author{Matthieu Wyart}
\affiliation{Center for Soft Matter Research, New York University, 4 Washington Place, New York, NY, 10003, USA}


\begin{abstract}
Connectedness and applied stress strongly affect elasticity in solids. In various amorphous materials, mechanical stability can be lost either by reducing connectedness or by increasing pressure.  We present an effective medium theory of elasticity that extends previous approaches by incorporating the effect of compression, of amplitude $e$, allowing one to describe quantitative features of sound propagation, transport, the boson peak, and elastic moduli near the elastic instability occurring at a compression $e_c$. The theory disentangles several frequencies characterizing the vibrational spectrum:  the onset frequency $\omega_0\sim \sqrt{e_c-e}$ where strongly-scattered modes appear in the vibrational spectrum, the pressure-independent frequency $\omega_*$ where the density of states displays a plateau, the boson peak frequency $\omega_{BP}$ found to scale as $\omega_{BP}\sim\sqrt{\omega_0\omega_*}$, and the Ioffe-Regel frequency $\omega_{IR}$ where scattering length and wavelength become equal.  We predict that sound attenuation crosses over from $\omega^4$ to $\omega^2$ behavior at $\omega_0$, consistent with observations in glasses. We predict that a frequency-dependent length scale $l_s(\omega)$ and speed of sound $\nu(\omega)$ characterize vibrational modes, and could be extracted from scattering data. One key result is the prediction of a flat diffusivity above $\omega_0$, in agreement with previously unexplained observations. We find that the shear modulus does not vanish at the elastic instability, but drops by a factor of 2. We check our predictions in packings of soft particles and study the case of covalent networks and silica, for which we predict $\omega_{IR} \approx \omega_{BP}$. Overall, our approach unifies sound attenuation, transport and length scales entering elasticity in a single framework where disorder is not the main parameter controlling the boson peak, in agreement with observations. This framework leads to a phase diagram where various glasses can be placed, connecting microscopic structure to vibrational properties.
\end{abstract}

\maketitle

\section{Introduction}

From granular materials and foams to molecular glasses and colloids, a wide range of amorphous materials exhibit a  transition from liquid-like to solid-like behavior. In the solid phase, these materials display anomalous elastic properties. In particular, amorphous solids universally present an excess of vibrational modes over the Debye model (that predicts a quadratic dependence of the density of vibrational modes with frequency), a phenomenon referred to as the `boson peak' \cite{phillips_book}. 
Phonon dispersion is observed to change sharply in the vicinity of the boson peak frequency: phase velocity displays a minimum, and sound attenuation changes its frequency dependence from $\omega^4$ to $\omega^2$ \cite{Baldi:2010,Baldi:2011b}. Thermal conductivity measurements support that above these intermediate frequencies modes are strongly scattered, and that their diffusivity (the frequency-dependent diffusion coefficient associated to heat transport) is small and independent of frequency \cite{Kittel:1949}, as observed numerically in packings of repulsive particles \cite{Vitelli:2010,Xu:2009}. These observations  are not understood, since a comprehensive theory of transport in amorphous solids is lacking. Moreover, they indicate the presence of at least one characteristic frequency scale, and through the sound speed a characteristic length scale, whose relation to disorder, however, remains controversial \cite{MonacoGiordano:2009,MonacoMossa:2009,Xu:2009}. As is well known, the static structure does not  indicate any obvious characteristic length scale larger than particle size \cite{Tanguy02,Leonforte06}. 

Beyond its importance for elasticity and transport, the boson peak relates to key features of the dynamics near the glass transition. In fragile liquids (for which the activation energy grows under cooling), the boson peak frequency decreases toward zero under heating, while its amplitude increases \cite{Tao:1991, chumakov:2004}. This observation has been interpreted \cite{Grigera:2002,Parisi:2003} as the existence of an elastic instability at some temperature $T^*$ where the boson peak frequency would vanish. At higher temperature, typical configurations are saddles, with many unstable directions in phase space. At lower temperature, a typical configuration lies near an energy minimum and vibrational modes are stable, a scenario already proposed by Goldstein \cite{goldstein}. Interestingly, the shear modulus increases rapidly under cooling in fragile liquids \cite{Dyre:2006, torchinsky}, an effect that could be responsible for most of the growth of the activation energy \cite{Dyre:2006}. One possibility is that the rapid change of the shear modulus stems from the proximity of elastic instability, and that the material stiffens as it is cooled past $T^*$ \cite{Wyart:2010b,Berthier:2011b}. However, predictions for the behavior of the shear modulus near an elastic instability are contradictory, as some predict that it should vanish at the instability \cite{YoshinoMezard:2010,Sheinman:2012}, while others predict that it does not \cite{Schirmacher:2007,Yoshino12}.

For these reasons, it is important to understand the nature of the boson peak, its associated length scales, and its relationship with elastic moduli. In most existing theories the presence of a peak results from disorder. More specifically, the boson peak has been proposed to emerge as a consequence of localized modes \cite{Buchenau:1992}, microscopic disorder in force constants \cite{TaraskinElliott:2002,Schirmacher98}, mesoscopic disorder in shear modulus \cite{Schirmacher:2006,Schirmacher:2007,Marruzzo13,Ferrante13}, properties of disordered matrices \cite{Grigera:2002,Grigera:2003}, or anharmonicity \cite{Gurevich:2003}. Some of these approaches, in particular \cite{Schirmacher:2006,Schirmacher:2007,Ferrante13}, can reproduce the $\omega^4$ to $\omega^2$ cross-over of the sound attenuation and the presence of a minimum in the speed of sound, but currently do not explain the flat diffusivity above this cross-over frequency. 
Most importantly, although disorder certainly affects sound dispersion and mode diffusivity, there is ample evidence that in many materials disorder is secondary in controlling the density of vibrational modes \cite{Chumakov:2014}.  This fact is well-established even in very disordered structures, such as random packings of particles \cite{Liu:2010} or silica \cite{dove,Wyart053}. It must be more generally true in the various materials where the boson peak is similar in the glass and in the crystal \cite{Chumakov:2011,Chumakov:2014}. 

Why in many materials does disorder strongly affect transport, but have such little effect on the density of vibrational modes? If it is not disorder, what in the microscopic structure controls the boson peak?  In recent years these questions have been addressed in simple amorphous solids made of repulsive short-range particles \cite{Liu:2010,OHern:2003, Silbert05, Wyart:2005a,Wyart:2005b,Wyart053}, in colloidal glasses \cite{BritoWyart:2009,BritoWyart:2006} and in covalent networks  \cite{Wyart053,dove,Yan:2013}. One central result \cite{Wyart:2005a,Wyart:2005b,Wyart053} is the stability diagram of Fig.\,\ref{fig:phasediagram}a, showing that in these systems the key microscopic parameters controlling mechanical stability are the coordination $z$ (the average number of contacts per particle in packings, or the valence in covalent networks), and the applied compressive strain $e \sim f/(kr)$ where $f,k,r$ are the typical force, stiffness, and distance between strongly interacting particles (a third important factor is the presence of weak interactions, such as Van der Waals interactions in covalent networks or long-range interactions in a Lennard-Jones, but this effect simply renormalizes the value of $e$, see below).  Physically this diagram implies that under compression, more contacts need to be formed to guarantee mechanical stability. On the line separating stable and unstable configurations, the boson peak frequency vanishes and its amplitude becomes very large. An important result of \cite{Wyart:2005a,Wyart:2005b} is that this phase diagram holds true independent of the amount of disorder, and thus applies to crystals as well. Thus if a crystal and a glass have similar local order, then they should have a similar boson peak amplitude. This is the case, for example, between silica and crystobalite \cite{dove,Wyart053}, but not so for radial short-range interactions, since in the latter case crystalline packings are much more coordinated than random ones \cite{Liu:2010}. 

\begin{figure}[ht!] 
\begin{tikzpicture}[scale=0.95]
  \fill[fill=blue!20!white] 
      (1,0) parabola (4,2.5) |- (0,0);
  \draw (1.2,1.5) node[above] 
      {$\mathrm{unstable}$};
  \draw (2.7,0.5) node[right] 
      {$\mathrm{stable}$};
  \draw (1,0) node[below] {$z_c$};

  \draw[->] (-0.2,0) -- (4,0) node[below] {$z$};
  \draw[->] (0,-0.2) -- (0,2.5) node[below left] {$e$};

  \draw (1,0) parabola bend (1,0) (4,2.5); 

\draw(0,2.7) node[above] {(a)};
\node[above right] at (4.2,-0.65){\includegraphics[viewport=15 20 360 330,width=0.23\textwidth]{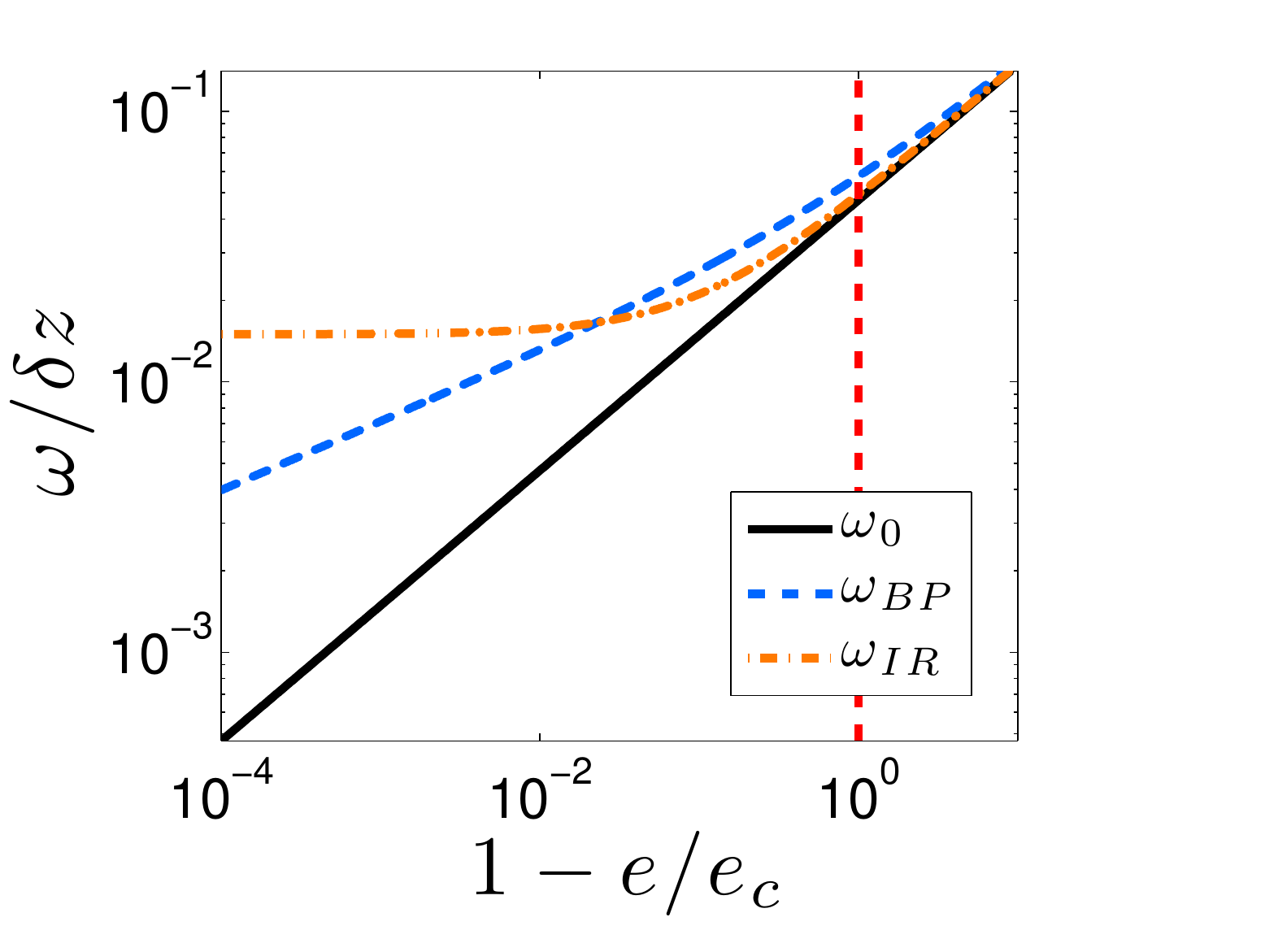}};
\draw(4.3,2.7) node[above] {(b)};
\end{tikzpicture}
\caption{\label{fig:phasediagram} (a) Stability diagram for packings of repulsive particles and elastic networks, where $e$ is compressive contact strain and $z$ is coordination. Stability requires $z \geq z_c$ and $e < e_c$, where $e_c(z) \sim (z-z_c)^2$. (b) Characteristic frequencies versus distance to elastic instability $1-e/e_c$, at small $\dz\geq0$. The onset frequency $\omega_0\sim \sqrt{e_c-e}$ is where strongly-scattered modes appear in the vibrational spectrum; $\omega_{BP}$ is the boson peak frequency; and $\omega_{IR}$ is the Ioffe-Regel frequency where scattering length and wavelength become equal. The density of states displays a plateau at $\omega_*$, such that for $e\geq0$ we have $\omega_* \sim \dz$. The vertical red line indicates $e=0$. For $e \ll e_c$ or $e\leq0$, all the frequencies are nearly identical, thus the spectrum is characterized by a single frequency scale. }
\end{figure}
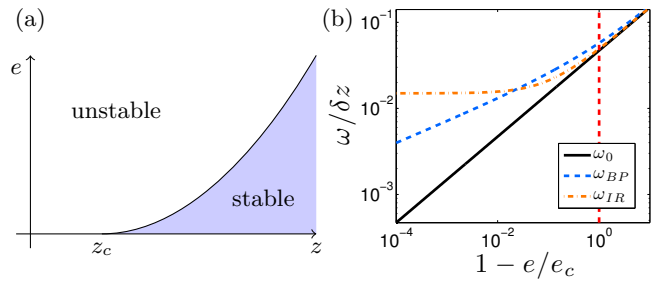

\begin{figure}[ht!] 
\centering
\begin{tikzpicture}[scale=0.95]
  \shade[top color=red!75!white,bottom color=blue!15!white,shading angle = 30,fill opacity=0.6] (0,0) rectangle (6,2.5);
  \shade[top color=red!45!white,bottom color=blue!15!white,shading angle = 5,fill opacity=0.6] (0,0) rectangle (6,2.5);
  \fill[fill=white] (0,0) parabola (4,2.52) |- (4,2.52) -- (0,2.52);
   \draw[red!45!white, ultra thick] (0,0) parabola bend (0,0) (4,2.5);
   \draw[red!35!white, ultra thick] (0.2,0) parabola bend (0.1,0) (4.1,2.5);
    \draw[dashed, ultra thick] (0.1,0) parabola bend (0.1,0) (4.3,2.5);
\draw (3.4,1.5) node[rotate=40,below right] {packings};
\draw (2.9,0.4) node[rotate=40,below right] {(emulsions, colloids)};

  \draw [fill=blue!15!white] (-3,-2) rectangle (6,0);
  \fill[fill=green!15!white] (0,0) parabola bend (-0.1,0) (-3,-1.6) -- (-3,0);
    
  \fill[fill=blue!15!white] (0,-0.8) circle [radius=0.2];
   \draw[dashed,ultra thick] (0,-0.8) circle [radius=0.2];
 \draw (-0.2,-0.7) node[left] {silica};
        
 \draw[dashed,ultra thick]  (-3,-1.2) --(3,-1.2) node[above] {chalcogenides} -- (6,-1.2);
 
   \draw[green!85!black, thick] (0,0) parabola bend (-0.1,0) (-3,-1.6);
   \draw[green!55!white, thick] (0,0) parabola bend (-0.2,0) (-3,-1.55);


  \draw[->] (-3,0) -- (6,0);
  \draw(5.5,0) node[below] {$z-z_c$};
  \draw[->] (0,-2) -- (0,2.5) node[below left] {$e$};

\end{tikzpicture}
\caption{\label{fig:phasediagram2} (Color) Schematic placement of amorphous solids (dashed lines) in stability diagram, where $e$ includes the effect of weak interactions, as discussed in the main text. In the red region ($1-e/e_c \ll 1$), vibrational properties are characterized by several distinct frequency scales, as shown in Figure \ref{fig:phasediagram}b, and proximity to elastic instability strongly affects transport. In the blue region, there is a single frequency scale. In the green region, there is a gap in the density of vibrational states at intermediate frequency. No solids can lie in the white region, which is unstable.}
\end{figure}
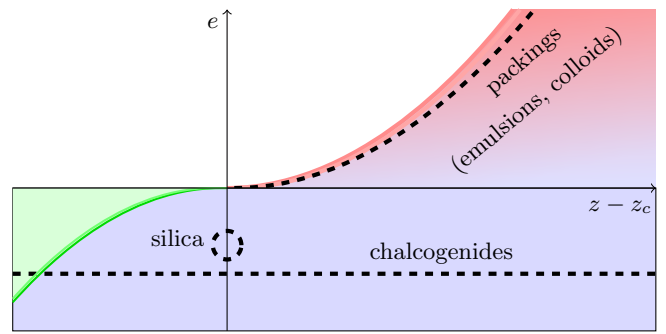

At the theoretical level, two approaches exist to compute vibrational properties in these systems. The phase diagram of Fig. \ref{fig:phasediagram}a was first derived with variational arguments \cite{Wyart:2005a,Wyart:2005b}
that apply independently of disorder, which allows one to predict the vibrational spectrum but is not informative on transport properties. To capture the latter, effective medium \cite{Kirkpatrick:1973, thorpe, Garboczi86,Schirmacher98,Choy:1999, Mao:2010, Wyart:2010, During:2013}, a self-consistent method based on a perturbation in the disorder amplitude, can be used when modes are not localized. Thorpe  \cite{thorpe,Garboczi86} applied this method to show that the shear modulus vanishes continuously near rigidity percolation, and Schirmacher \cite{Schirmacher98} argued that this transition is discontinuous when interactions with negative stiffnesses are included. 

Recently it was shown that this approach  captures quantitatively the singularity of some vibrational properties  of repulsive particles near the unjamming transition \cite{Wyart:2010}, where the coordination reaches the Maxwell threshold $z_c=2d$ where $d$ is the spatial dimension. In particular, the density of vibrational modes and its characteristic frequency $\omega_*\sim \delta z \equiv z-z_c$ \cite{Silbert05,Wyart:2005b}, the mode diffusivity \cite{Vitelli:2010,Xu:2009} and the length scale $l_c\sim 1/\sqrt{\delta z}$ characterizing the modes at the boson peak  \cite{Silbert05} are reproduced.  However this calculation  assumed that {\it no applied stress is present}, i.e. the axis $e=0$ in Fig.\ref{fig:phasediagram}, an approximation that certainly breaks down for repulsive particles, but applies to elastic networks with weak spatial fluctuations of coordination \cite{Wyartmaha}. Moreover, (i)  the sound attenuation was not considered in \cite{Wyart:2010} forbidding a comparison with scattering data  and (ii) the role of applied pressure on transport, on the shape of the density of vibrational modes, on length scales and on elastic moduli was not derived. Understanding the effect of compression is particularly relevant for packings of particles and colloidal glasses, as these systems lie very close to the stability line of Fig. \ref{fig:phasediagram}a \cite{BritoWyart:2006,BritoWyart:2009,Wyart:2005b}, implying that the effect of pressure in these systems is very strong. Moreover, elastic instabilities in supercooled liquids are expected to generically occur at $e\neq0$. (iii) The proposed framework allows one to  classify vibrational and transport properties in various glasses, such as silica and covalent networks, based on their structure.   As we will see, this comparison is rich and non-trivial. We will argue that the two-parameter theory of linear vibrational properties in amorphous solids we propose, while still reasonably simple, is necessary to obtain a framework unifying observations in systems as different as covalent networks and colloidal glasses. 

 In this work we extend the effective medium approximation to describe at a microscopic level systems under  compression, where contacts carry a force. Although we provide a simplified description where all contacts have the same stiffness, our formalism is readily extendable to heterogeneous contacts \cite{DeGiuli14}. Our simplified description can, however, capture  the presence of weak interactions. 
 Our central results are: 
 \begin{enumerate}
\item Our effective medium approximation captures the phase diagram of Fig. \ref{fig:phasediagram}a. At a compressive strain $e_c\sim (z-z_c)^2$ an instability occurs. 
\item The shear modulus remains finite at elastic instability, and simply decreases by a factor of 2 as $e$ is increased toward $e_c$.
\item We can compute four frequencies: the onset frequency $\omega_0\sim \sqrt{e_c-e}$ where strongly-scattered modes appear in the vibrational spectrum, the pressure-independent  frequency $\omega_*$ where the density of states  displays a plateau, the  boson peak frequency  $\omega_{BP}\sim\sqrt{\omega_0\omega_*}$, and the Ioffe-Regel frequency $\omega_{IR}\sim \omega_*$ where scattering length and wavelength become equal. These four frequencies are nearly identical only for $e\ll e_c$ or negative $e$, and display three distinct scalings as $e\to e_c$, as shown in Figure \ref{fig:phasediagram}b. 
\item  The sound attenuation $\Gamma(\omega) \sim \omega^4$ for $\omega<\omega_0$ and $\Gamma(\omega) \sim \omega^2$ for $\omega_0 < \omega < \omega_*$.
\item The speed of sound is minimal at $\omega_0$.
\item Our analysis indicates that to infer transport properties like diffusivity from scattering data, it is more convenient to analyze the dynamical structure factor at fixed $\omega$ rather than at fixed wave number $q$. This 
approach allows one to compute a frequency-dependent speed of sound $\nu(\omega)$ and scattering length $\ell_s(\omega)$. We argue that above the boson peak, these quantities differ significantly from the approximation used in the literature to extract them. In the intermediate and high frequency regime, capturing correctly these quantities is important to describe transport. Their scaling with frequency is predicted.
\item We build a theory of transport that applies to non-localized modes. In particular we find that the mode diffusivity does not depend on frequency as soon as the density of states deviates from the Debye behavior (i.e. for  $\omega>\omega_0$), in agreement with previous numerical observations in sphere packings \cite{Vitelli:2010,Xu:2009}.
\item  The length scale below which continuum elasticity breaks down is $\ell_c\sim \omega_0^{-1/2}$, as shown in a companion paper \cite{Lerner:2013b}.
 \end{enumerate}
   Results 2,4,5 have been previously obtained in different models, see e.g. \cite{Kohler13,Marruzzo13}. These approaches however  assume that the boson peak stems from spatial fluctuations in elasticity, at odds with our work.  

Finally, we compare these predictions to experimental and numerical observations in glasses and particle packings, where many of our scaling results agree with observations.  We discuss where certain glasses, such as silica, chalcogenides, colloids and soft particles are placed in our phase diagram, allowing us to make predictions on their transport properties.
 Overall, our approach unifies sound attenuation, transport,  elastic length scales (discussed in a companion paper) and the boson peak in a framework where disorder is secondary in controlling the peak amplitude, in agreement with observations in many materials. 
 
\section{Model}
\begin{figure}[ht!] 
\begin{tikzpicture}[scale=0.95]
\draw(0,4.1) node[above] {(a)};
\draw(4.5,4.1) node[above] {(b)};
\node[above right] at (-0.5,-0.2){\includegraphics[viewport = 10 10 450 210,clip,width=0.47\textwidth]{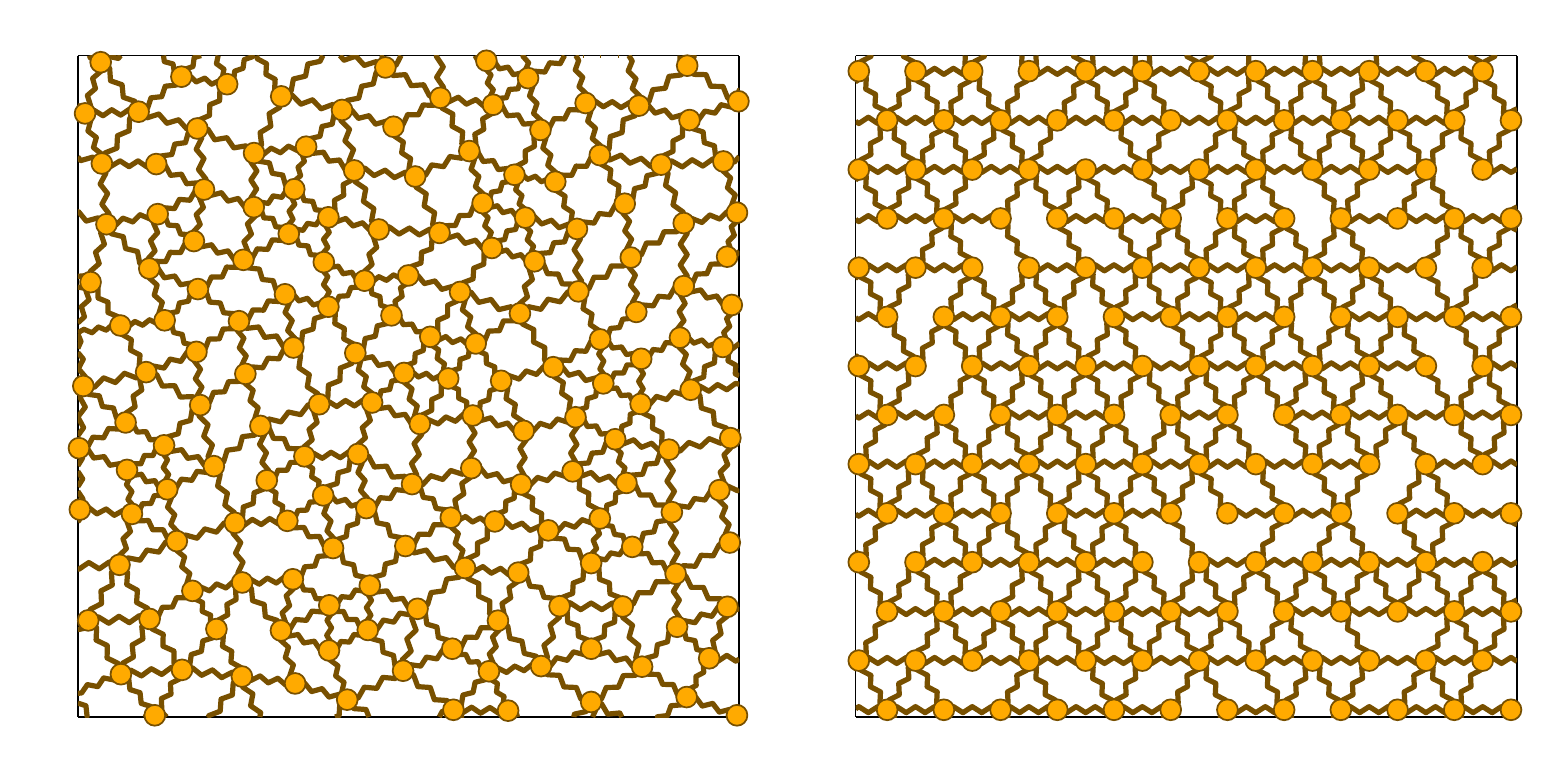}};
\end{tikzpicture}
\caption{\label{fig:lattice} (a) Illustrative random network with small fluctuations in coordination, and (b) diluted regular lattice. Note that these illustrations are in 2D, but the theory is constructed in 3D.}
\end{figure}

{\it Ingredients to be incorporated:} We seek to compute how salient aspects of the microscopic structure of glasses affect their vibrational properties. We should focus on the following features, that have been argued to control the boson peak  in a variety of materials \cite{Wyart:2005b,Wyart053,Xu:2007,Liu:2010}:

(i) The connectedness $z$, or more precisely the excess connectedness $\delta z\equiv z-z_c$ with respect to the minimal connectedness  $z_c$ required for rigidity. The notion that structures must be sufficiently connected to be mechanically stable is fundamental in engineering since the work of Maxwell \cite{Maxwell:1864}. For an elastic network, for example as shown in Figure \ref{fig:lattice}a, the connectedness  is simply the coordination, i.e. the average number of springs per node. In a packing of purely repulsive, short-range particles, it is the average number of  contacts per particle.  For radial interactions in general, Maxwell showed that $z_c=2d$  \cite{Maxwell:1864}.  When interactions have a long-range component such as in a Lennard-Jones glass, a distinction must be made between strongly and weakly interacting particles \cite{Wyart053,Xu:2007}, which allows to define $z$ as the coordination of the  networks of strong interactions.  In general, the definition of connectedness depends on the system. For example, for generic covalent networks, $z$ is the valence; if elements of different valences are present, $z$ can be changed continuously by monitoring the composition. For such multi-body interactions one finds $z_c=2.4$ \cite{Phillips}.

(ii) The compressive strain $e$. It is well known that an applied pressure can lead to elastic instability, such as the buckling of thin rods and shells.  It is also true in a bulk solid. To see this, consider two interacting particles, forming a contact $\alpha$. If they are displaced relative to each other,
the expansion of the energy is to second order \cite{LandauLifshitz:1986,Alexander:1998}  
\eq{
\delta E_\alpha = \frac{k_\alpha}{2} | r_\alpha^\parallel |^2 - \frac{f_\alpha}{2 \sigma_\alpha} | r_\alpha^\perp |^2, \label{dE}
}
where $| r_\alpha^\parallel |$ is the norm of the longitudinal displacement (in the direction of the contact), $| r_\alpha^\perp |$ is the norm of the perpendicular (transverse) displacement, $f_{\alpha}$ is the force in the bond $\alpha$ (by convention, positive if the force is repulsive), $ k_\alpha$ is the stiffness of the interaction, and $\sigma_\alpha$ is the distance between the two particles. 
Since any longitudinal displacement increases $\delta E_\alpha$, the longitudinal term is stabilizing. However, if the contact is under compression, $f_\alpha >0$ and the transverse term is \textit{destabilizing}: this is a geometrical consequence of the fact that any small transverse displacement at $\alpha$ necessarily increases the center-to-center distance between the particles, and therefore lowers the energy, if the interaction is repulsive. The last term in Eq.(\ref{dE}) can be considered a spring orthogonal to the contact, of stiffness $-f_\alpha/\sigma_\alpha$. The relative contribution of the transverse to the longitudinal term is characterized by a dimensionless number $e_\alpha=\frac{f_\alpha}{k_\alpha \sigma_\alpha}$, whose typical value in the material is denoted $e$, which is positive under compression. 
As we show below, the role of pre-stress $e$ can be important in amorphous solids even when $e\ll 1$  \cite{Wyart:2005b}, if the excess coordination is small ($z\approx z_c$ or smaller). 

(iii) If there is a hierarchy in the strength of the interactions involved (for example in covalent networks the Van der Waals interactions are much weaker than covalent bonds), it is useful to introduce a dimensionless number $W=k_{weak}/k_{strong}$ where $k_{weak}$ ($k_{strong}$) is the characteristic stiffness of the weak (strong) interaction \cite{Wyart053,Xu:2007,Yan:2013}. In our model, we do not explicitly consider this possibility. However, weak interactions play a role very similar to the transverse stiffness induced by negative forces in the contacts. Both perturbations are negligible, except  if the network of strong interactions is not well-connected ($z\approx z_c$ or smaller). The transverse {\it vs} longitudinal aspect of these perturbations is not expected to make a qualitative difference. Accordingly, weak interactions effectively renormalize the value of $e$, decreasing it by some amount proportional to $W$: the system is stabilized by weak interactions. 

%

{\it Model:} Elastic networks are arguably the simplest models to study the role of connectedness and applied stress on vibrational properties in amorphous structures. 
However, different geometries of networks can be considered. In rigidity percolation,  bonds are diluted randomly from a lattice \cite{thorpe2}, as shown in Figure \ref{fig:lattice}b.
This model has the disadvantage that large fluctuations of coordination appear, so that near the rigidity transition the rigid (mechanically stable) cluster is a fractal object that excludes a finite fraction of bonds. 
This scenario does not apply to granular and colloidal systems \cite{Liu:2010} and is not believed to occur in covalent glasses either \cite{Boolchand}, because large fluctuations of coordination or density are penalized energetically. 
In this model computing elastic  properties near the rigidity transition remains a challenge, and mean field methods such as effective medium give incorrect results \cite{thorpe2}. 
A better model is constructed by removing the large spatial fluctuations of coordination while keeping the network random,  as illustrated in Fig.\ref{fig:lattice}a; this can be done in several ways and leads to results very similar to packings of particles \cite{Wyartmaha}.  It was recently found that the elastic properties in such networks can be computed accurately, at least in the absence of pressure. In particular when effective medium -- a mean-field approximation that neglects large spatial fluctuations of coordination-- is used on the rigidity percolation model --where fluctuations are important -- analytical results describe accurately elastic networks with weak spatial fluctuations of coordination \cite{Wyart:2010}. In this work, we use the same strategy of using jointly the rigidity percolation model (under an applied pressure) together with effective medium. As we will see, this procedure allows to make accurate predictions that are verified in amorphous solids, such as particle packings.

In practice, we consider an isotropic lattice of coordination $z_0 > z_c$ in three dimensions, which is randomly diluted to reach a final coordination $z=z_c+\dz$; an illustrative example of a diluted lattice is shown in Figure \ref{fig:lattice}b. 
We will take $z_0=12$ \footnote{ $z_0$ is a parameter of the theory, which connects to coordination fluctuations. Indeed, random bond dilution implies a relationship between the fluctuations in coordination and $z_0$. At fixed $z$, larger $z_0$ implies larger coordination fluctuations. If $v$ is the variance of particle coordination, then $v=z(1-z/z_0)$ for random independent dilution.}. To model random dilution and compression of the lattice, each spring constant and contact force are set to nonzero values $k_\alpha = k_0$ and $f_\alpha = e k_0 \sigma_\alpha$ with probability $P=z/z_0$, and 0 with probability $1-P$, independently at each bond. We make the approximation that the force in each bond is identical (which violates force balance as soon as $P<1$, but nevertheless leads to accurate predictions, see below).  
We also assume  that springs are weakly deformed $\sigma_\alpha=\sigma$, which is asymptotically valid when $e \ll 1$.  Note that in this model  the compressive strain $e$ and pressure $p$ are linearly related, but this is not exact in general \footnote{For our model $p =  e \rho z k_0 \sigma/(2d)$, with $\rho=N/V$ the number density. However, in a Lennard-Jones glass for example, where attraction is weak and long-range in comparison with repulsion, the relevant microscopic parameter $e=\langle f_\alpha/(k_\alpha \sigma_\alpha) \rangle$ can be nonzero even at $p=0$: in that case the long-range attraction contributes a tension equal and opposite to the repulsion, but at different typical bond lengths, say $\sigma_{att}$ and $\sigma_{rep}$. The relative contribution of tension to strain is smaller by a factor $\sim (\sigma_{rep}/\sigma_{att})^2 \ll 1$ so that $e>0$.}. 

\section{Effective Medium} 
 
We study the average effect of compression and coordination when our diluted lattice is forced at frequency $\omega$. We write $|\delta \bm{R} \rangle=(\delta \Rb_1, \delta \Rb_2, \ldots, \delta \Rb_N)$ for the vector of node displacements and $|\bm{F} \rangle = (\bm{F}_1, \bm{F}_2, \ldots, \bm{F}_N)$ for the vector of applied forces. Then an imposed oscillatory force $|\bm{F}\rangle  e^{i \omega t}$ causes a response 
\eq{ \label{Gb}
|\delta \bm{R} \rangle = \Gb(\omega) |\bm{F}\rangle e^{i \omega t},
}
where $\Gb(\omega)$ is the Green's function, a $dN \times dN$ matrix, defined precisely in Appendix A. As discussed below, all vibrational properties can be written in terms of $\Gb$. 

The Green's function $\Gb(\omega)$ appearing in \eqref{Gb} will depend on the particular realization of the random geometry, for example the location of absent springs from bond dilution (see Figure \ref{fig:lattice}b). To compute average vibrational properties, we seek its disorder average $\Gbo(\omega)$. Effective medium theory (EMT) is an approximation scheme that is well-suited to describing disorder-averaged properties, such as $\Gbo(\omega)$ \cite{Choy:1999, Mao:2010, Wyart:2010, Sheinman:2012, During:2013}. In EMT, the random diluted lattice is approximated by an effective regular lattice with identical effective stiffnesses, which are functions of $\omega$, $e$ and $\dz$. 
The effective stiffnesses are chosen uniquely by the following physical requirement. One considers the lattice where a single bond $\alpha$, located at the origin,  is assumed to be disordered (in our case, it is present with probability $P$) while all the other bonds have the same effective stiffnesses. We demand that the disorder average (on the bond at the origin) of the true Green's function, $\Gbo$, should be equal to the effective Green's function $\Gb_{E}$. 
How this is done in practice is shown in Appendix A.

 Because longitudinal and transverse displacements play a different role in \eqref{dE},  our EMT has both longitudinal and transverse effective spring constants, $\tilde \kpa$ and $\tilde \kpe$, respectively. $\tilde \kpe$ can be thought as a spring orthogonal to the contact, which captures that orthogonal displacements have a finite stiffness when contact forces are non-zero. For notational convenience, we drop the tilde of $\tilde \kpa$ and introduce  $-e  \kpe\equiv \tilde \kpe$. In effective medium, $\kpa$ and $\kpe$ depend on frequency $\omega$ and, in general become complex, capturing the fact that vibrational modes scatter off disorder and decay.  
 

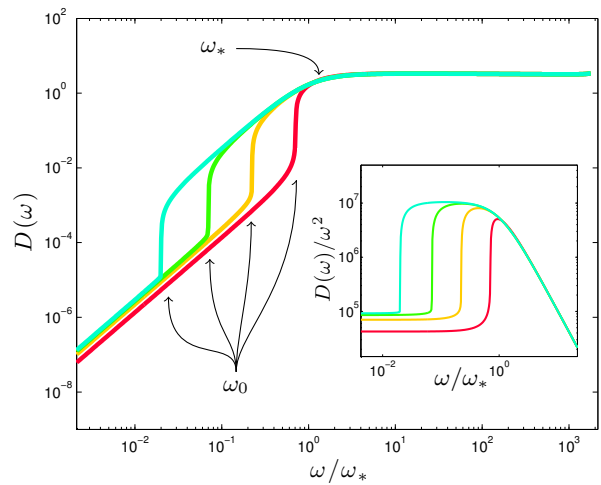
\begin{figure}[ht!] 
\begin{tikzpicture}[scale=1]
\clip (0,0) rectangle (8,6.5);
\node[above right]{\includegraphics[viewport=20 7 460 360,width=0.49\textwidth]{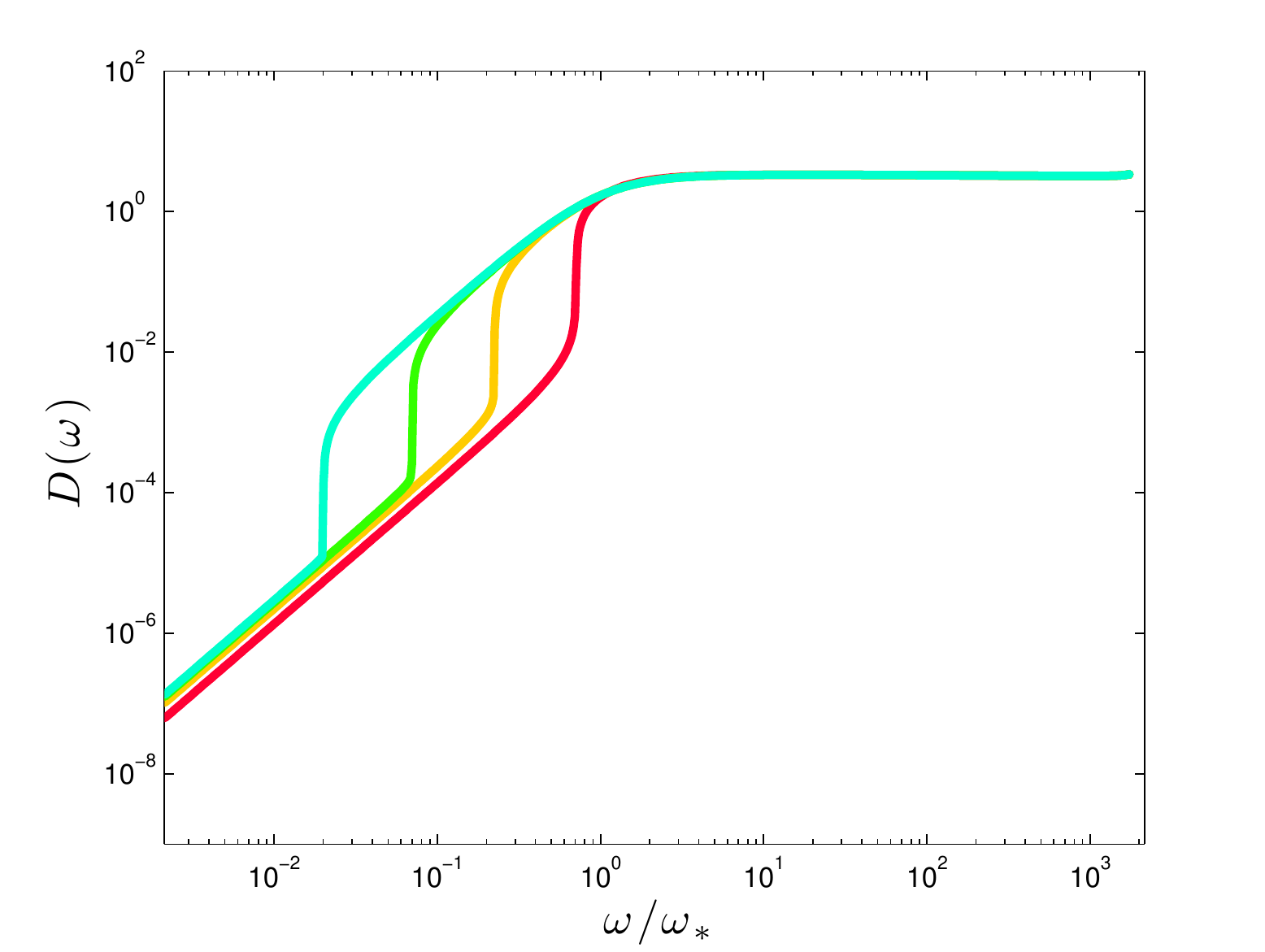}};
\draw[->] (3,1.5) .. controls (3,2) and (2.1,1.5) .. (2.1,2.5);
\draw[->] (3,1.5) .. controls (3,2) and (2.65,2.0) .. (2.65,3.0);
\draw[->] (3,1.5) .. controls (3,2) and (3.2,2.6) .. (3.2,3.50);
\draw[->] (3,1.5) .. controls (3,2) and (3.8,2.6) .. (3.8,4.05);
\draw(3,1.5) node[below] {$\omega_0$};
\draw[->] (3,5.8) .. controls (4.0,5.8) and (4.1,5.6) .. (4.1,5.45);
\draw(3,5.8) node[left] {$\omega_*$};
\node[above right] at (3.70,1.3){\includegraphics[viewport=5 7 410 360,width=0.21\textwidth,clip]{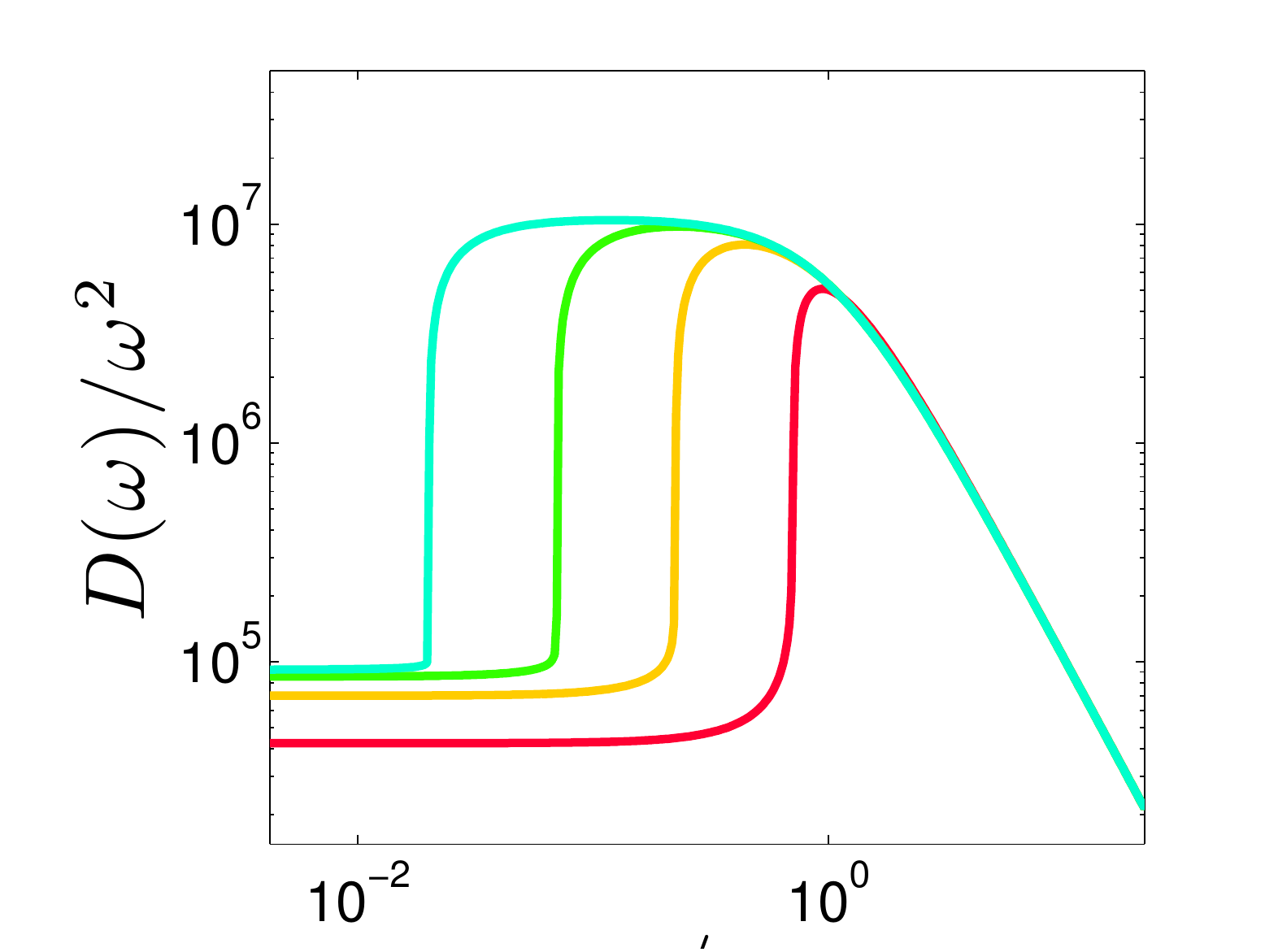}};
\node at (6,1.4){$\omega/\omega_*$};
\end{tikzpicture}
\caption{\label{fig:Domega} Density of states $D(\omega)$ as strain approaches its critical value, at $\dz=0.012$. From left to right, the distance to instability is $1-e/e_c=0.0005,0.005,0.05,0.5$ (blue,green,orange,red online). Inset: Reduced density of states $D(\omega)/\omega^2$ vs $\omega/\omega_*$, showing a boson peak at $\omega_{BP} \sim \sqrt{\omega_0 \omega_*}$.}
\end{figure}

\begin{figure}[ht!] 
\begin{tikzpicture}[scale=1]
\clip (0,0) rectangle (8,6.5);
\node[above right]{\includegraphics[viewport=25 7 460 360,width=0.49\textwidth]{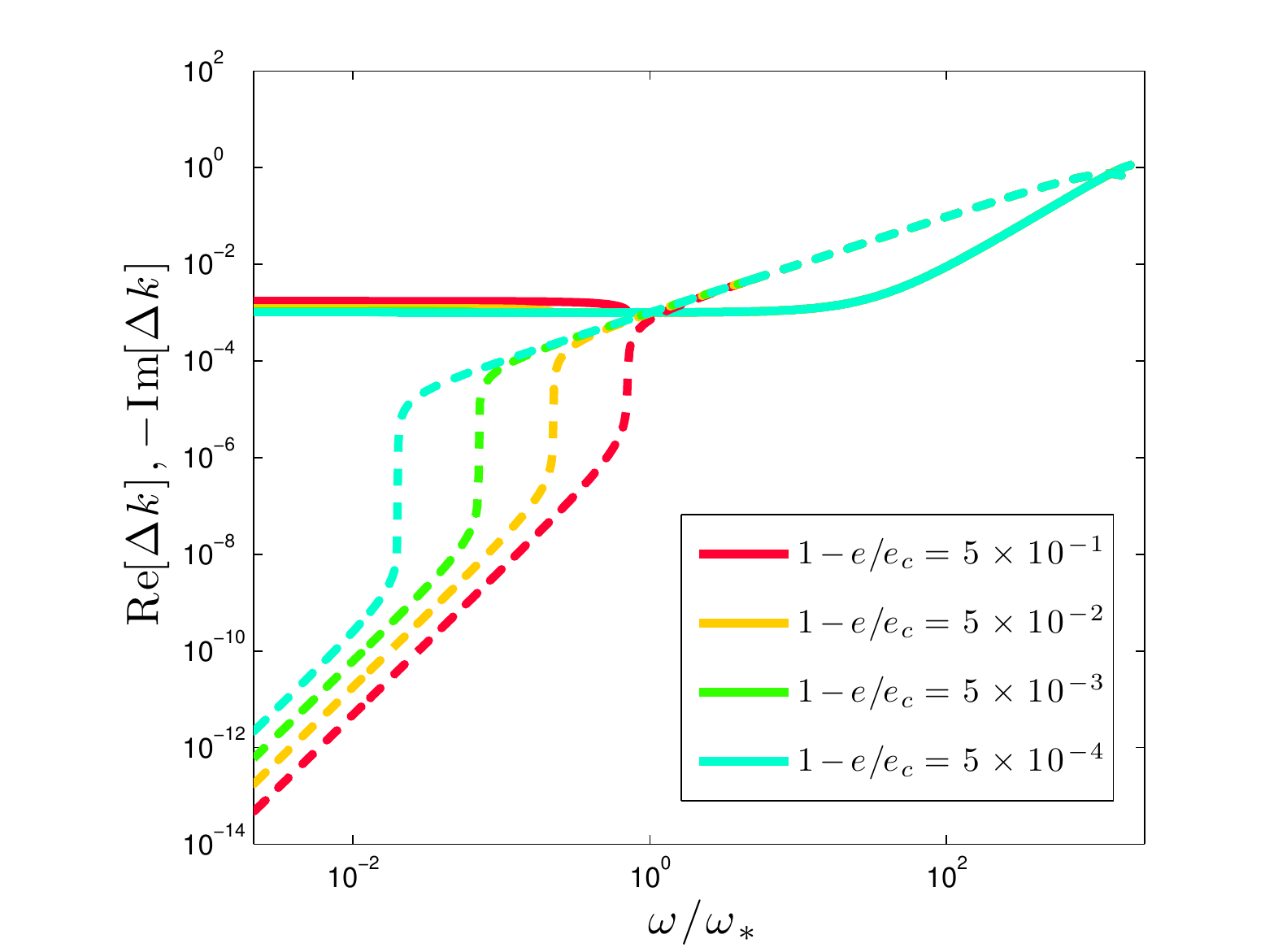}};
\draw[->] (3,1.5) .. controls (3,2) and (2.45,2.6) .. (2.45,3);
\draw[->] (3,1.5) .. controls (3,2) and (3.1,2.7) .. (3.1,3.1);
\draw[->] (3,1.5) .. controls (3,2) and (3.6,2.8) .. (3.6,3.2);
\draw[->] (3,1.5) .. controls (3,2) and (4.1,3.0) .. (4.1,3.4);
\draw(3,1.5) node[below] {$\omega_0$};
\draw[->] (3.6,5.6) .. controls (3.8,5.6) and (4.3,5.2) .. (4.3,4.7);
\draw(3.6,5.6) node[left] {$\omega_*$};
\end{tikzpicture}
\caption{\label{fig:shearmod} Real (solid) and minus-imaginary (dashed) parts of complex shear modulus $\Delta k$ as strain approaches its critical value, at $\dz=0.012$. Colours are as in Figure \ref{fig:Domega}.  }
\end{figure}
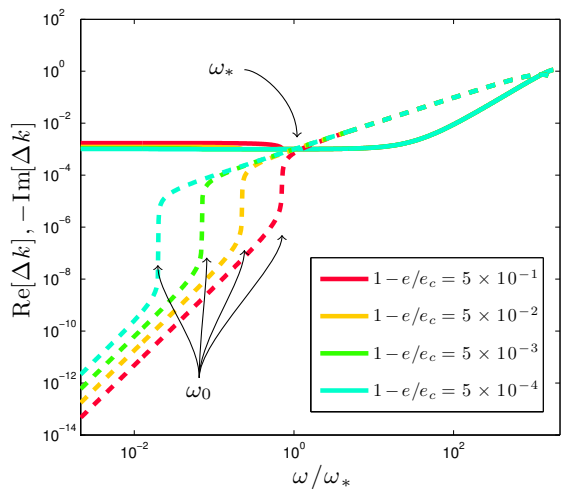



 
Using standard EMT techniques, discussed in Appendix A, we derive simple algebraic equations for $\kpa$ and $\kpe$ in terms of $\Gbo$. For simplicity, we neglect the difference between the longitudinal and transverse speed of sound. Since we are interested in low-frequency behaviour, for the Green's function we consider
\eq{ \label{G2}
\overline{\Gb}(\rb,\omega) = \frac{z_0}{d} \delb \int_{BZ} \frac{d^d q}{(2\pi)^d} \frac{e^{i \qb \cdot \rb}}{(\kpa-e \kpe) q^2 - m \omega^2},
}
where $BZ = \{ \qb: \; |\qb| < \Lambda  \}$ is an approximate first Brillouin zone. This is the continuum Green's function for an elastic medium with shear modulus $\kpa - e \kpe$, equal longitudinal and transverse sound velocities, cutoff at a (dimensionless) microscopic wavenumber $\Lambda$ and appropriately renormalized.    


Equation \eqref{G2} and EMT equations \eqref{emt1}, \eqref{emt2}, and \eqref{G1} in Appendix A define a closed system for $\kpa$ and $\kpe$, which we solve numerically and analytically in the limit $\dz \to 0$, with $d=3$, $z_0=12$, and $\Lambda=\pi$. Below we  focus on the case $e>0$ and $\delta z\geq 0$ relevant for repulsive spheres and colloids, and come back to the case $e<0$ when we discuss covalent networks and silica. We take units with $m$, $\sigma$, and the bare stiffness $k_0$ equal to unity.

\section{Results}
The EMT gives an expression for the complex shear modulus $\Delta k=\kpa-e\kpe$, with which we obtain the effective Green's function $\Gbo$, and all derived quantities.

\subsection{Density of States}

The density of vibrational states $D(\omega)$ is determined using the identity $D(\omega)=(2\omega/\pi) \mbox{ Im[tr[}\overline{\Gb}(0,\omega)]]$ and plotted in Figure \ref{fig:Domega}. 
For small $\delta z$ and frequencies $\omega \lesssim \dz$, we can solve these equations analytically (see Appendix B), and find:
\eq{ \label{Domega}
D(\omega) = \begin{cases} C_3 \frac{\omega^2}{(\omega_*+\sqrt{\omega_0^2-\omega^2})^{3/2}} & \mbox{ if } \omega < \omega_0  \\
C_4 \frac{\omega \sqrt{\omega^2-\omega_0^2}}{\omega^2 + \omega_*^2 - \omega_0^2} & \mbox{ if } \omega > \omega_0, \end{cases}
}
where the frequency scales are
\eq{
\omega_* & = c_1 \dz, \\
\label{ee} \omega_0 & = c_2 \sqrt{e_c-e},
}
as announced above. The positive constants $c_i$ and $C_i$ are non-universal: in our framework they depend on $z_0$ and $\Lambda$. However, the exponents associated with $\omega_*$, $\omega_0$, and $e_c$ are independent of these microscopic details. 

Elastic instability occurs when an eigenvalue $\omega^2$ becomes negative. This occurs when $\omega_0=0$, hence $e_c$ is the critical strain. It scales as
\eq{
e_c(\dz) = (c_1/c_2)^2 \dz^2,
}
in agreement with Figure \ref{fig:phasediagram}a. This implies $\omega_0 < \omega_*$. As we show in detail below, in general, there are three regimes: a Debye regime $\omega<\omega_0$ in which the solid behaves like an elastic continuum, a high-frequency regime $\omega > \omega_*$ in which $D(\omega)$ has a plateau, and an intermediate regime $\omega_0 < \omega < \omega_*$. Asymptotically, 
\eq{
D(\omega) \sim \begin{cases} \omega^2/\dz^{3/2} & \mbox{ if } \omega \ll \omega_0 \notag \\
\omega^2/\dz^2 & \mbox{ if } \omega_0 \ll \omega \ll \omega_* \notag \\
1 & \mbox{ if } \omega \gtrsim \omega_*  \end{cases}
}
The boson peak frequency is conventionally defined by the maximum of $D(\omega)/\omega^2$. We find
\eq{
\omega_{BP} & \approx \half \sqrt{\omega_0} \sqrt{3 \omega_0 + \sqrt{8\omega_*^2 + \omega_0^2}} \notag \\
& \sim \sqrt{\omega_0 \omega_*},
}
which is between $\omega_0$ and $\omega_*$: see Figure \ref{fig:phasediagram}b and inset to Figure \ref{fig:Domega}. When $\omega_0 \ll \omega_*$, its amplitude scales as $D(\omega_{BP})/\omega_{BP}^2 \sim \omega_*^2$.



\subsection{Elastic Modulus}

The complex shear modulus $\Delta k=\kpa-e\kpe$ is plotted in Figure \ref{fig:shearmod}. Its behaviour is captured by the first terms in an asymptotic solution, 
\eq{\label{deltak}
\Delta k(\omega,e) & = C_1 \omega_* + C_1 \sqrt{\omega_0^2 - \omega^2} \\
& \qquad - \frac{i C_2\omega^3}{\sqrt{\omega_0^2-\omega^2} \sqrt{\omega_* + \sqrt{\omega_0^2-\omega^2}}} + \curlyO(\dz^{2}), \notag
}
where $\sqrt{\omega_0^2-\omega^2} = -i \sqrt{\omega^2-\omega_0^2}$ for $\omega> \omega_0$. 

The static shear modulus is $\mu=\mbox{Re }\Delta k(\omega=0)$. We find
\eq{ \label{mu}
\mu = C_1 \big(\omega_* + \omega_0\big) \sim \dz \big(1 + \sqrt{1-e/e_c}\big).
}
We predict that $\mu$ remains finite at elastic instability, reduced by a factor of 2 from its unstressed value.

\subsection{Sound Dispersion}

Sound dispersion at frequency $\omega$ is determined by the large $r$ behaviour of $\overline{\Gb}(\rb,\omega)$. For $r \gg 1$, we find \footnote{The inverse Fourier transform can be done with the method of steepest descent and the Residue Theorem. The constant $C_5=z_0/(12\pi)$.}  
\eq{ \label{Gdecay}
\overline{\Gb}(\rb,\omega) \sim C_5 \frac{1}{\Delta k} \frac{1}{r} \delb \; e^{i \omega r/\nu(\omega)} e^{-r/l_s(\omega)},
}  
with scattering length
\eq{ \label{ls}
l_s(\omega) = \frac{1}{\omega} \frac{|\Delta k|}{|\mbox{Im }\Delta k^{1/2}|}
}
and phonon speed
\eq{ \label{nu}
\nu(\omega) =\frac{|\Delta k|}{\mbox{Re }\Delta k^{1/2}}. 
}
 With Eq. \eqref{deltak}, we can use these equations to determine explicitly the scaling of the relevant scattering length and phonon speed at frequency $\omega$.  

\begin{figure}[ht!] 
\begin{tikzpicture}[scale=1]
\clip (0,0) rectangle (8,6.5);
\node[above right]{\includegraphics[viewport=20 7 460 360,width=0.49\textwidth]{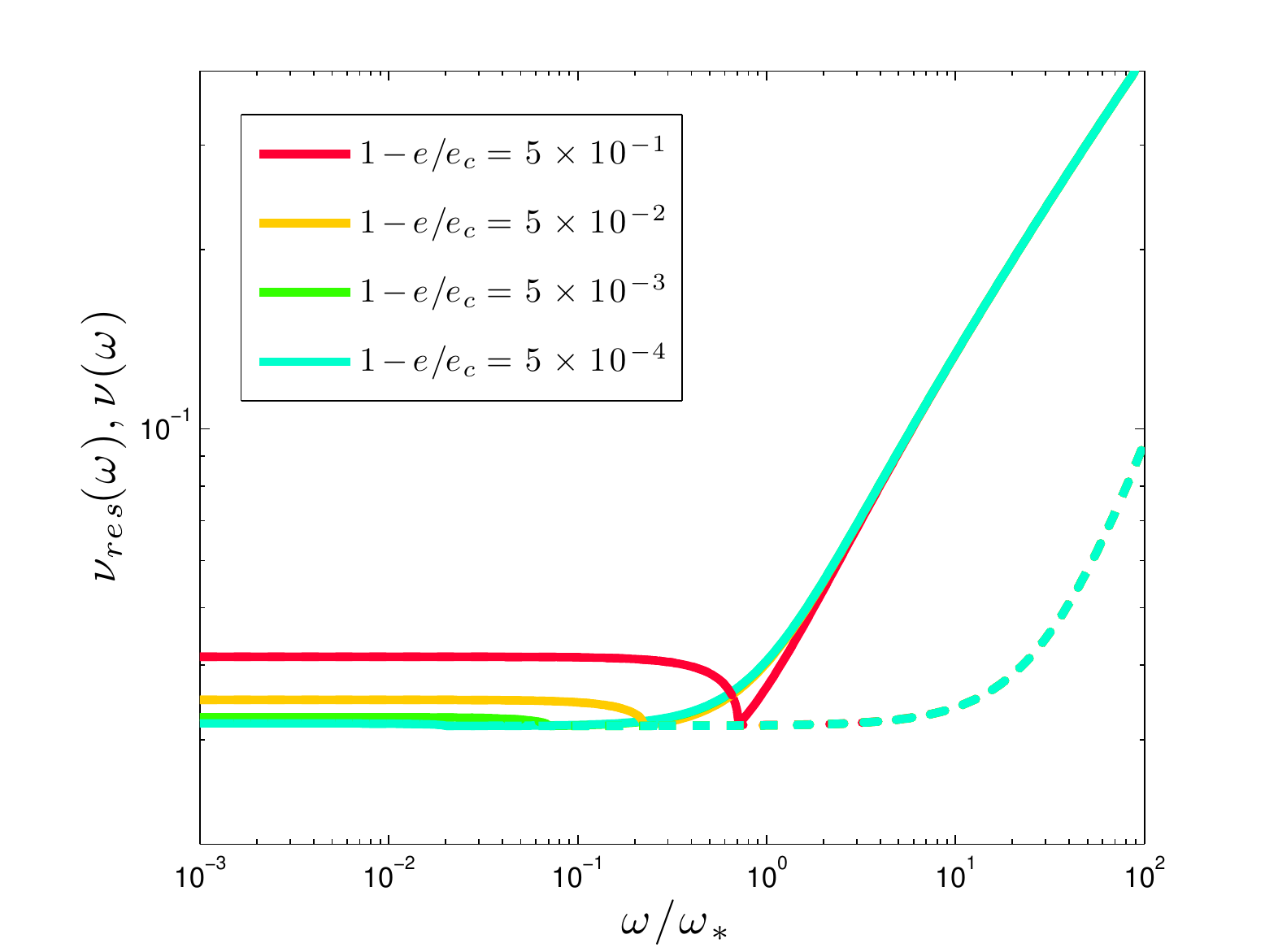}};
\draw[->] (3,3.3) .. controls (5.0,3.3) and (5.5,3.3) .. (5.9,3.4);
\draw(3,3.3) node[left] {$\nu$};
\draw[->] (3,2.6) .. controls (5.0,2.6) and (6.5,2.3) .. (7.0,2.2);
\draw(3,2.6) node[left] {$\nu_{res}$};
\end{tikzpicture}
\caption{\label{fig:nu} Phonon speed $\nu$ (solid) and resonant wave speed $\nu_{res}$ (dashed) as strain approaches its critical value, at $\dz=0.012$. The curves are indistinguishable below $\omega_{IR}$.}\end{figure}
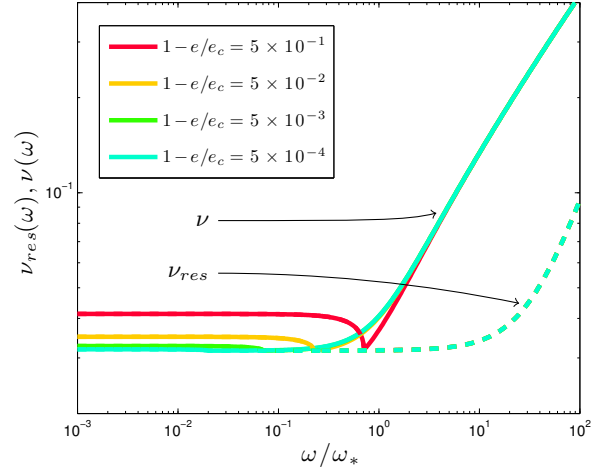

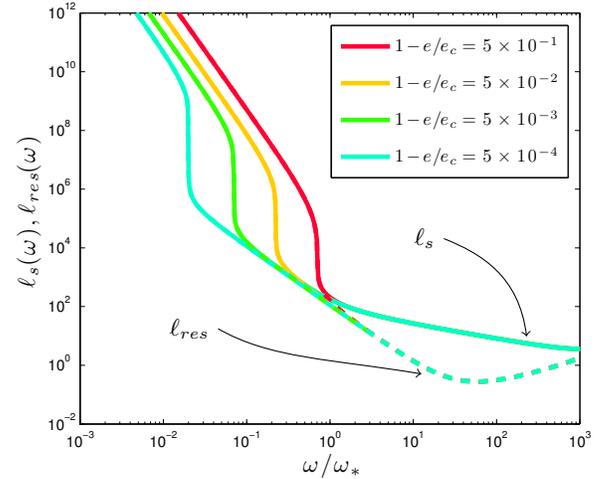
\begin{figure}[ht!] 
\begin{tikzpicture}[scale=1]
\clip (0,0) rectangle (8,6.5);
\node[above right]{\includegraphics[viewport=20 7 460 360,width=0.49\textwidth]{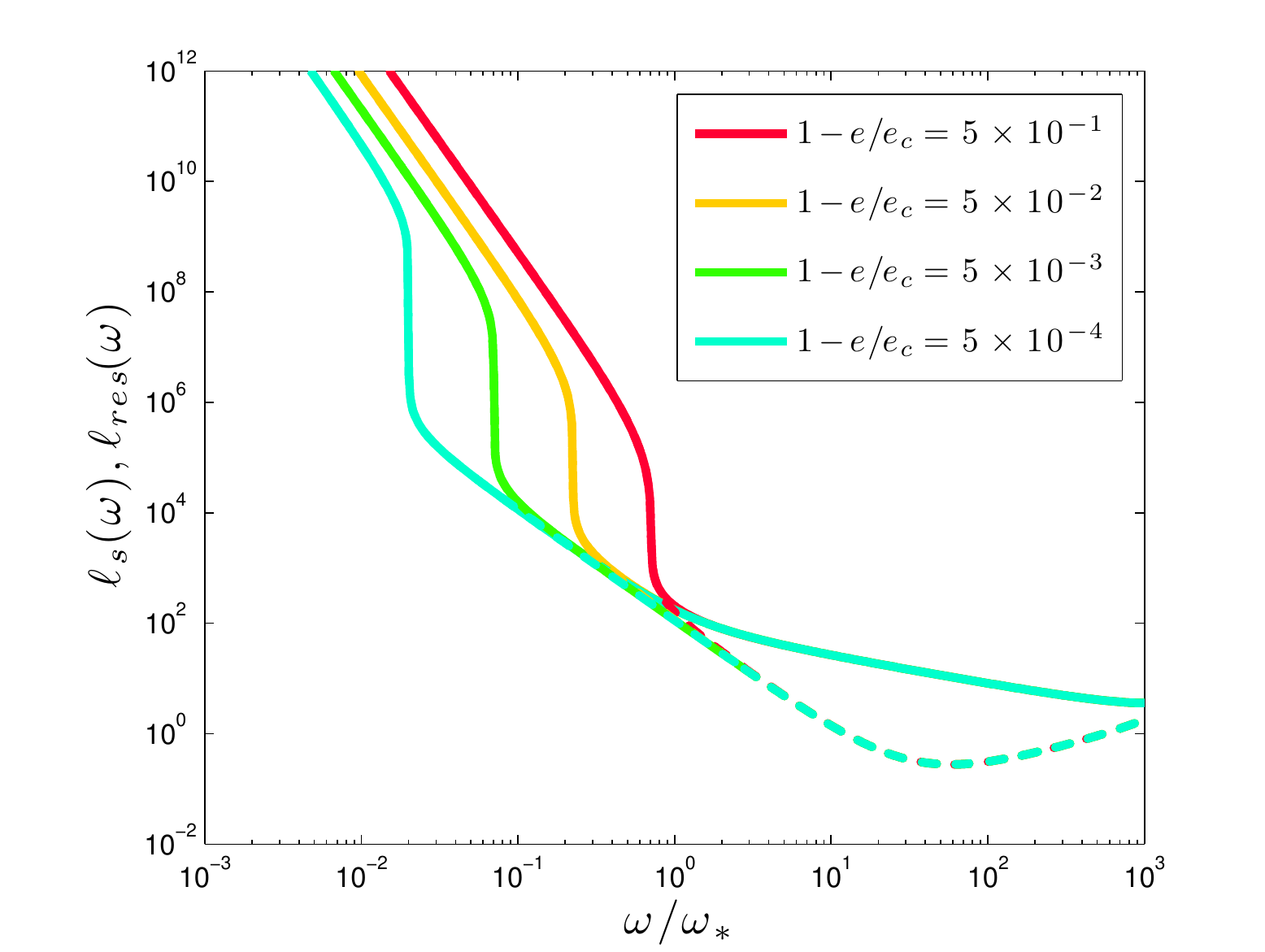}};
\draw[->] (3,2) .. controls (3.7,1.6) and (4.5,1.6) .. (5.7,1.4);
\draw(3,2) node[left] {$\ell_{res}$};
\draw[->] (6,3.2) .. controls (6.5,3.0) and (7.0,2.7) .. (7.1,1.9);
\draw(6,3.2) node[left] {$\ell_s$};
\end{tikzpicture}
\caption{\label{fig:ell} Scattering length $\ell_s$ (solid) and resonant phonon scattering length $\ell_{res}$ (dashed) as strain approaches its critical value, at $\dz=0.012$. The curves are indistinguishable below $\omega_{IR}$.}\end{figure}

We find that phonon speed $\nu$ has a minimum at $\omega_0$, as shown in Figure \ref{fig:nu}. Asymptotically,
\eq{
\nu(\omega) \sim \begin{cases} 
\sqrt{\dz} & \mbox{if} \; \omega \ll \omega_* \\ 
(\omega^2+\omega_*^2)^{1/4} & \mbox{if} \; \omega \sim \omega_*
\end{cases}
}
The scattering length $\ell_s$ displays Rayleigh scattering for $\omega < \omega_0$, a sudden drop at $\omega_0$, and anomalous scattering above $\omega_0$, shown in Figure \ref{fig:ell}. Asymptotically, 
\eq{
\ell_s(\omega) \sim \begin{cases} 
\omega^{-4} \dz^3 & \mbox{if} \; \omega < \omega_0 \\ 
\big(\omega \sqrt{\omega^2-\omega_0^2}\big)^{-1}  \dz^{3/2} & \mbox{if} \; \omega_0 < \omega \ll \omega_* \\ 
\omega^{-1} (\omega^2+\omega_*^2)^{1/4} & \mbox{if} \; \omega \sim \omega_*
\end{cases}
}
The relevance of the scattering length $\ell_s$
for the breakdown of continuum elasticity is discussed in a companion paper \cite{Lerner:2013b}.


{\it Transport}: The identification of physically-relevant length and velocity scales at frequency $\omega$ strongly constrains the frequency dependence of sound dispersion properties. Indeed, the only dimensionless parameter that can be formed from $\omega, \nu(\omega),$ and $\ell_s(\omega)$ is $n(\omega)=\ell_s \omega/\nu$; physically, $2\pi n$ is the number of wavelengths the response at frequency $\omega$ travels before scattering. 


\begin{figure}[ht!] 
\begin{tikzpicture}[scale=1]
\clip (0,0) rectangle (8,6.5);
\node[above right]{\includegraphics[viewport=20 7 460 360,width=0.49\textwidth]{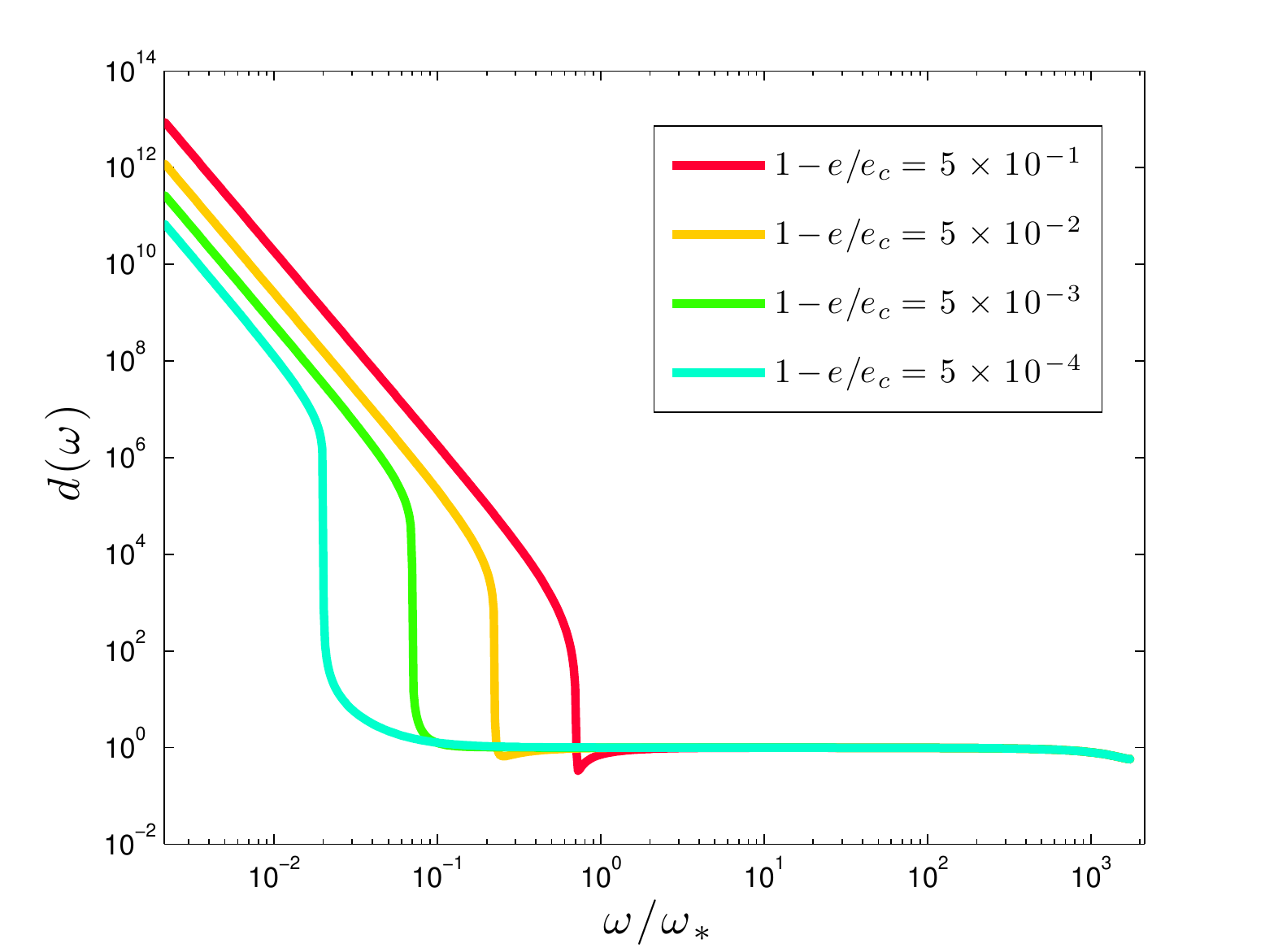}};
\end{tikzpicture}
\caption{\label{fig:diff} Energy diffusivity $d(\omega)$ as strain approaches its critical value, at $\dz=0.012$. } 
\end{figure}

Energy transport by phonons is characterized by their energy diffusivity, $d(\omega)$ \cite{Vitelli:2010,Xu:2009}. In Appendix C we use Kubo formulae to calculate $d(\omega)$ within the effective medium approximation, using crucially the asymptotic behaviour of $\Gbo$, equation \eqref{Gdecay}. The result is
\eq{ \label{domega}
d(\omega) \approx \ell_s(\omega) \nu(\omega) f[n(\omega),\ell_s(\omega)],
}
where 
\eq{ \label{f}
f(n,\ell_s) = C_6 \frac{4n^2}{(n^2+1)^2} + C_7 \frac{n^2+1}{n^2-1 +\pi \Lambda \ell_s}.
}
The factor $f[n(\omega),\ell_s(\omega)]$ tends to a constant both at large and small $\omega$; in particular, we have $f \approx C_7$ for $\omega < \omega_0$, and $f \approx C_6$ for $\omega \gtrsim \omega_*$. In the intermediate regime $\omega_0 < \omega < \omega_*$, $f$ exhibits nontrivial behaviour. These results have a simple physical interpretation: for $\omega<\omega_0$ and $\omega > \omega_*$, the diffusivity is accurately estimated on dimensional grounds as $\propto \ell_s \nu$, the natural diffusion constant at frequency $\omega$. This gives Rayleigh scattering $d \sim \omega^{-4}$ in the Debye regime, and a plateau $d \sim 1$ for $\omega > \omega_*$, as argued in \cite{Wyart:2010}. In the intermediate regime, modes are hybrids of plane waves and `anomalous modes' that appear above $\omega_*$ \cite{Wyart:2005a}, and the diffusivity has nontrivial $n$ dependence. 
Our central new result is that the diffusivity is predicted to be flat all the way down to $\omega_0$, as shown in Figure \ref{fig:diff}, as observed numerically \cite{Xu:2009,Vitelli:2010}.  




\section{Thermal Conductivity}

Thermal conductivity $\kappa(T)$ can be calculated from energy diffusivity using\cite{Schirmacher:2006}
\eq{ \label{kappa}
\kappa(T) \propto \int d\omega \; D(\omega) d(\omega) \frac{\omega^2}{T^2} \frac{e^{\hbar \omega/k_B T}}{(e^{\hbar \omega/k_B T}-1)^2}.
} 
In real glasses, as $\omega \to 0$, $d(\omega)$ transitions from Rayleigh scattering $\sim\omega^{-4}$ to an anharmonic regime where phonons scatter on two-level systems, not accounted for in our harmonic expression \eqref{domega}. However, the high $T$ behaviour of $\kappa(T)$ is expected to be unaffected by this anharmonicity. Here our aim is simply to prove that the flat density of states and diffusivity we predict at high frequency can capture well the high temperature behavior of the thermal conductivity, in agreement with an early observation by Kittel. To show this, we cutoff the integral in \eqref{kappa} below frequencies $\omega_0/2$, where the diffusivity rises from its plateau. 

\begin{figure}[ht!] 
\begin{tikzpicture}[scale=1]
\clip (0,0) rectangle (8,6.5);
\node[above right]{\includegraphics[viewport=20 7 460 360,width=0.49\textwidth]{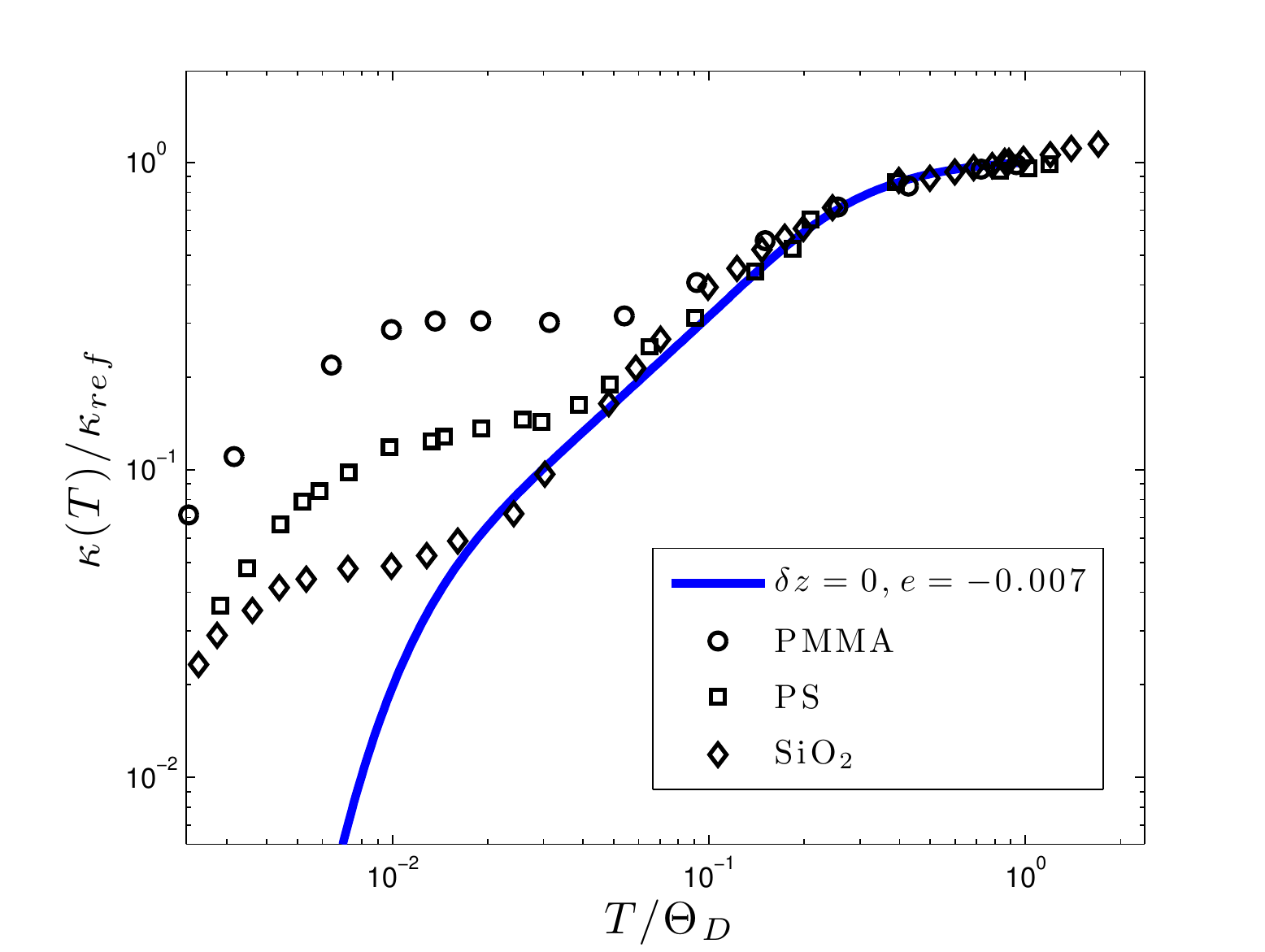}};
\end{tikzpicture}
\caption{\label{fig:kappa} Thermal conductivity $\kappa(T)$ from theory (solid) at indicated values of $e$ and $\dz$, and data from Freeman and Anderson \cite{FreemanAnderson:1986} (symbols) on (polymeric) PMMA ($\bigcirc$), PS ($\square$), and SiO$_2$ ($\Diamond$), in arbitrary units.} 
\end{figure}

The result is plotted in Figure \ref{fig:kappa}, where it is compared with data from Freeman and Anderson \cite{FreemanAnderson:1986} on vitreous PMMA, PS and SiO$_2$. For realistic values of $\dz$ and $e$ (expected for silica, as discussed in the comparison section below) we can quantitatively capture the data above the plateau of thermal conductivity. Note that our prediction works for a larger range of temperature for silica than for other materials.  In our view, this  reflects the fact that silica is nearly isostatic and thus displays a flat diffusivity over a large frequency range. 

%

\section{Comparison with Scattering Experiments}

The Green's function $\Gb(\omega)$ is not directly accessible in experiments on molecular glasses.  However, inelastic neutron and x-ray scattering experiments measure a derived quantity, the inelastic dynamic structure factor, $S_{in}(q,\omega)$. For harmonic dynamics, $S_{in}(q,\omega)=(k_B T/(d\pi)) \mbox{Im}[q^2 \omega^{-1} \tr(\Gb(q,\omega))]$ \cite{Baldi:2011a}; this leads to
\eq{ \label{Sin}
S_{in}(q,\omega) \propto k_B T \frac{-q^4 \omega^{-1} \mbox{Im}[\Delta k]}{(\omega^2 - q^2 \mbox{Re}[\Delta k] )^2 + (\mbox{Im}[\Delta k])^2 q^4 },
}
 a form consistent with earlier theory \cite{Schirmacher:2006,Schirmacher:2007}. 
A promising avenue to test the present theory  is to use the form \eqref{Sin}, with Im$[\Delta k(\omega)]$ and Re$[\Delta k(\omega)]$ treated as unknown functions, to fit scattering data. Using equations \eqref{ls} and \eqref{nu}, one can then obtain from $\Delta k(\omega)$ the scattering length $\ell_s$ and phonon speed $\nu$. 

The inelastic dynamic structure factor $S_{in}$ is not usually fitted to the form \eqref{Sin}, but instead to a damped harmonic oscillator form
\eq{ \label{Sindata}
S_{in}(q,\omega) \propto k_B T  \frac{q^2 \Gamma(q)}{(\omega^2-\Omega^2(q))^2+\omega^2 \Gamma^2(q)}.
}
In this expression, $\Omega(q)$ is the resonant frequency, and $\Gamma(q)$ is the full-width-half-maximum of the peak, known as the sound attentuation parameter \cite{ShintaniTanaka:2008,Baldi:2010,Ruta:2012}. The phase speed of the resonant mode is $\nu_{res}(q) = \Omega(q)/q$. By evaluating $\nu_{res}$ at the resonant wavenumber $Q=2\pi \Omega^{-1}(\omega)$, one obtains $\nu_{res}(\omega)$. Similarly, we let $\Gamma(\omega) \equiv \Gamma(Q(\omega))$. 

To compare with these fits, we can also define $\nu_{res}$ and $\Gamma$ in our theory (although in our theory these quantities are not the natural ones to consider at large frequencies). To do so, we note that when Im$^2[\Delta k] \ll \mbox{Re}^2[\Delta k]$, we can identify in \eqref{Sin} a resonant wavenumber $Q = \omega/\sqrt{\mbox{Re}[\Delta k(\omega)]}$; equivalently, the resonant frequency $\Omega$ satisfies $\sqrt{\mbox{Re}[\Delta k(\Omega)]}=\Omega/Q$, implying that the resonant phase velocity is $\nu_{res}(\omega)=\sqrt{\mbox{Re}[\Delta k(\omega)]}$. Its behaviour is shown in Figure \ref{fig:nu}: there is a minimum at $\omega_0$, and a very small increase up to $\omega_*$, the same range where the boson peak frequency is located. 

\begin{figure}[ht!] 
\begin{tikzpicture}[scale=1]
\clip (0,0) rectangle (8,6.5);
\node[above right]{\includegraphics[viewport=16 7 457 360,width=0.49\textwidth]{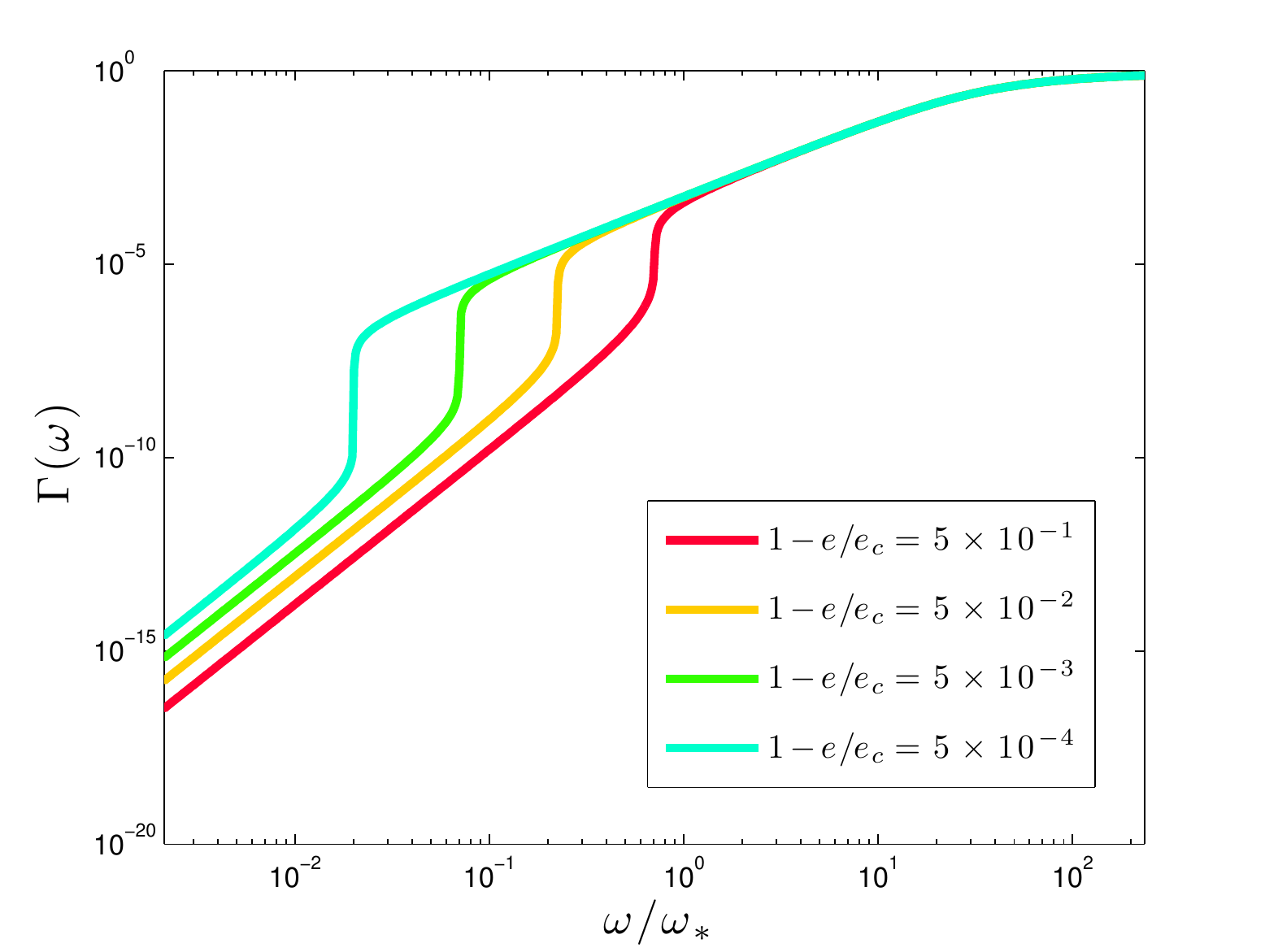}};
\end{tikzpicture}
\caption{\label{fig:Gammaomega} Sound attenuation $\Gamma(\omega)$ as strain approaches its critical value, at $\dz=0.012$. }
\end{figure}

The sound attenuation parameter $\Gamma(\omega)$ is the full-width-half-maximum of the peak; in our theory this is
\eq{
\Gamma(\omega) = - Q^2 \omega^{-1} \mbox{Im}[\Delta k(\omega)], 
}
plotted in Figure \ref{fig:Gammaomega}. We predict a transition from $\omega^4$ to $\omega^2$ at $\omega_0$, with a jump that increases in magnitude as the critical pressure is approached. 

In experiments on vitreous silica, $\nu_{res}$ is observed to have a minimum at the boson peak frequency, and $\Gamma(\omega)$ is observed to transition from $\omega^4$ below the boson peak frequency to $\omega^2$ above it \cite{Baldi:2010,Baldi:2011b}. Our theory is fully consistent with these results, if $\omega_{BP} \approx \omega_0$. Below  we will argue that  silica corresponds to $\delta z=0$, $e<0$ for which this property is satisfied, as illustrated in Figure \ref{fig:phasediagram}b. 

The interpretation of $S_{in}$ in terms of a resonant peak is not appropriate as soon as the scattering length of phonons is equal to half their wavelength. This defines the Ioffe-Regel frequency $\omega_{IR}$ at which $\pi\Gamma=\Omega$ \cite{ShintaniTanaka:2008}. We predict 
\eq{
\omega_{IR} \approx \sqrt{\omega_0^2 + \omega_*^2/\pi^2}
}
It has been suggested that the boson peak frequency is equal to $\omega_{IR}$. As illustrated in  Figure \ref{fig:phasediagram}b, we find that it depends on the microscopic structure of the glass. For sphere packings $e\rightarrow e_c$ (see below) and  $\omega_{IR}$ and $\omega_{BP}$ are predicted to differ.  For silica and well-coordinated covalent networks $e<0$ and we predict $\omega_{IR}\approx \omega_{BP}$, in agreement with observations \cite{Baldi:2010}.

Above $\omega_{IR}$, modes no longer resemble plane waves. This is apparent in Figure \ref{fig:nu}, which shows that $\nu \neq \nu_{res}$ above $\omega_{IR}$: the phase velocity of the total response is not characterized at all by $\nu_{res}$. Similarly, the scattering length of the resonant wavenumber, $\ell_{res}(\omega) = 2\nu_{res} \Gamma^{-1}$, differs from $\ell_s$ above $\omega_{IR}$, as shown in Figure \ref{fig:ell}. The difference can be dramatic: we see that for $\omega \sim 30\omega_*$, the total response persists for nearly 2 decades longer than the contribution from the resonant wavenumber. These comments underline the relevance of $\nu$ and $\ell_s$ as the physical velocity and length scales in the entire frequency range (below the localization transition).

\section{Comparison with specific glasses}

Our approach predicts that the vibrational properties of amorphous solids depend on their excess-coordination $\delta z=z-z_c$ and compressive strain $e$ (keeping in mind that the presence of weak interactions can be incorporated by lowering the value of $e$). In the $(\delta z, e)$ plane there is a forbidden region where no mechanically stable glasses are possible, represented in white in Fig. \ref{fig:phasediagram2}. Purely repulsive particles such as elastic spheres and colloidal glasses must lie in the red-blue region, which is stable despite $e>0$. In network glasses at small pressure, such as chalcogenides, $z-z_c$ can be monitored by changing the valence, allowing to explore the blue part of the phase diagram of Fig. \ref{fig:phasediagram2}. In this case $e<0$ due to the presence of weak Van der Waals interactions that can stabilize the system even if $z-z_c<0$.
We shall recall below why silica  corresponds to $\delta z=0$ and $e<0$. We now discuss the consequence of this classification in each case. 

{\it Repulsive short-range particles:} In granular materials, emulsions, and hard sphere colloidal glasses the particle interaction is repulsive and short-range, implying that $e>0$. Thus these systems  lie in the upper right corner of the phase diagram of Fig. \ref{fig:phasediagram2}. Considerable attention has been given in the ``jamming" literature to the case of frictionless spheres  interacting via a finite range potential \cite{Liu:2010}, because vibrational properties display critical properties as compression vanishes. To some extent, this scaling behavior can be experimentally observed \cite{Liu:2010}, in particular in emulsions (see e.g. \cite{brujic2}). Numerically, scaling exponents can be extracted precisely, which enables stringent testing of theories of elasticity and transport in amorphous materials. 

 Here we consider particles interacting via a one-sided harmonic potential (extension to other potentials, e.g. Hertzian, is straightforward). The contact compression is simply proportional to the increase of packing fraction $e\propto\phi-\phi_c$, where $\phi_c$ is the point at which pressure vanishes. Another useful system to consider can be made by replacing particles by points, and contacts by harmonic springs at rest \cite{Wyart:2005b}. This essentially removes the pre-stress term in the expansion of $\delta E$, see Eq. \eqref{dE}, and corresponds to setting $e=0$ in our formalism. It was found previously that (i) the shear modulus satisfies $\mu\sim \delta z$ independently of pre-stress \cite{OHern:2003,Wyartmaha}, and the prefactor of this relation decreases as the compressive pre-stress  increases \cite{Ellenbroek:2009,vanHecke:2010}. (ii)  $\delta z\sim \sqrt{e}$ \cite{OHern:2003}. (iii) The vibrational spectrum displays one frequency scale $\omega_*\sim \delta z$ above which $D(\omega)$ displays a plateau \cite{Wyart:2005b,Silbert05}. This is true independent of the pre-stress, but with pre-stress $D(\omega)$ is much larger for $\omega<\omega_*$ \cite{Wyart:2005b}, and presents non-plane-wave-like modes up to very small frequencies. (iv) The diffusivity is essentially independent of frequency in packings, but presents a behaviour consistent with Rayleigh scattering below $\omega_*$ when pre-stress is removed \cite{Vitelli:2010,Xu:2009}.

All these behaviours follow precisely our predictions, if we suppose that packings lie very close to the stability boundary in Figure \ref{fig:phasediagram}a.  Such marginal stability was proposed in \cite{Wyart:2005b} to rationalize the structure of packings; proposed explanations for this behaviour can be found in \cite{Wyart:2005b,BritoWyart:2009,charbonneau:2013}. Our prediction that $\mu$ does not vanish as $e\to e_c$ is thus important to understand why the shear modulus of packings is finite and scales as $\sim\dz$. Most importantly, being close to marginal stability, $e\approx e_c$, implies that $0\approx \omega_0 \ll \omega_{BP} \ll \omega_*\sim \delta z$, as illustrated in Fig. \ref{fig:phasediagram}. Since  we predict that the diffusivity presents a plateau above $\omega_0$, it must be independent of frequency in marginally stable packings, as indeed observed-- a key support to our theory of transport. When no pre-stress is present, our theory is consistent with the previous result of \cite{Wyart:2010} predicting a cross-over from Rayleigh scattering to flat diffusivity at $\omega_*\sim \delta z$.

\begin{figure*}[ht!] 
\begin{tikzpicture}[scale=1]
\node[above right]{\includegraphics[width=0.98\textwidth]{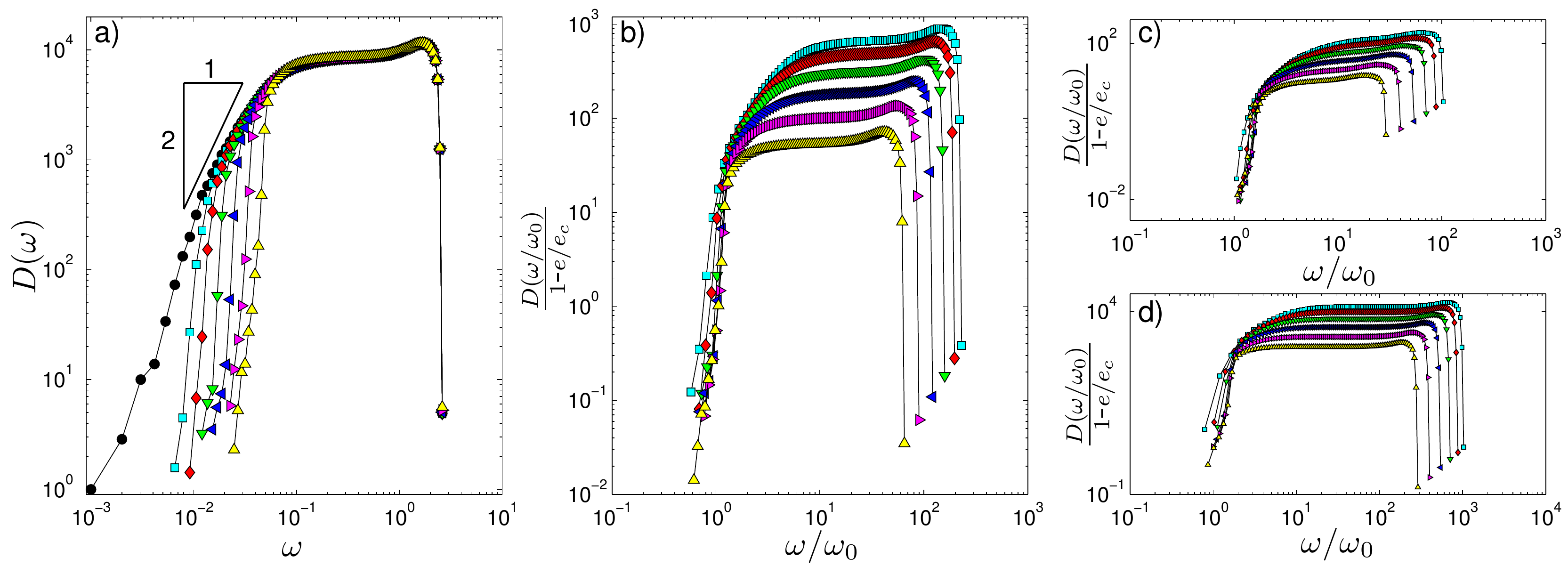}};
\end{tikzpicture}
\caption{\label{fig:Domegadata} (a) $D(\omega)$ from numerical packings in 3D at $e_0=10^{-3}$, whose contact forces have then been rescaled by a factor $1-x$, with $x=0,0.025,0.05,0.1,0.2,0.4,0.8$ (black,cyan,red,green,purple,yellow); rescaled $D(\omega/\omega_0)/(1-e(x)/e_c)$ at (b) $e_0=10^{-3}$, (c) $e_0=10^{-2}$, (d) $e_0=10^{-4}$. In all cases we find collapse of the onset frequency $\omega_0$ by assuming $e_0/e_c = 0.96$. Note that the original packing (black) has been omitted from the rescaled plots. The predicted slope in the intermediate regime, $2$, is shown in (a).}
\end{figure*}

\begin{figure}[ht!] 
\begin{tikzpicture}[scale=1]
\node[above right]{\includegraphics[width=0.45\textwidth]{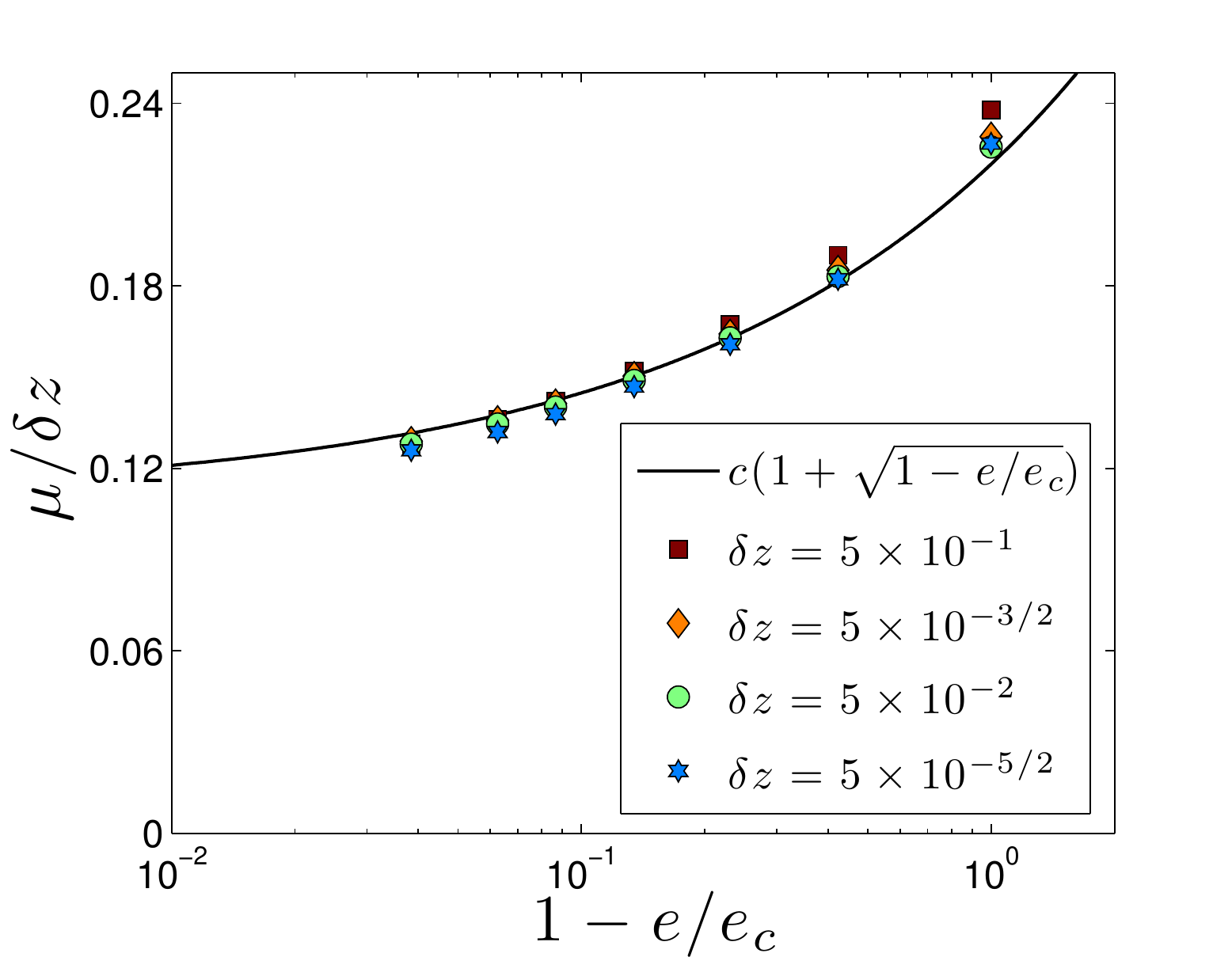}};
\end{tikzpicture}
\caption{\label{fig:mu} Shear modulus $\mu$ from numerical packings in 2D (symbols) at $\dz=5 \times 10^{-1}$ to $\dz=5 \times 10^{-{5/2}}$, for various $e(x)$, as discussed in the main text, and compared to theoretical prediction (solid). The shear modulus drops by a factor $\approx 2$ from $e=0$ to $e=e_c$. The constant $c=0.11$.}
\end{figure}

Both to test our scaling predictions for $\omega_0$ {\it vs} $e$, and to measure precisely how close sphere packings are from an elastic instability,  we construct 4000 bidisperse packings (size ratio $=1.4$) of 2000 frictionless spheres using the FIRE algorithm \cite{FIRE}, at strains $e_0=10^{-2},10^{-3}$, and $10^{-4}$. From each packing, we then manually rescale all contact forces by a factor $1-x<1$ and compute the resulting density of states $D_x(\omega)$, shown in Figure \ref{fig:Domegadata}a. If our packings were exactly at marginal stability with strain $e_0=e_c$, then we would have $e(x)=e_c (1-x)$, and predict a frequency scale $\omega_0\propto x^{1/2}$. However, this frequency does not collapse our data. 
Instead, as shown in Figures \ref{fig:Domegadata}b,c,d, we find a satisfactory collapse of all the data by assuming that packings are very close, but at a finite distance from an elastic instability, with $e_0/e_c = 0.96$ for \textit{all 3 pressures}. We have also rescaled the vertical axis, which collapses well the density of states in the regime $\omega_0 < \omega \ll \omega_*$ as it should according to Eq.\eqref{Domega}. The companion paper \cite{Lerner:2013b} presents further evidence that the distance to marginal stability in our packings  is 4\%, and checks that this number is not system-size-dependent.
Finally, the theory further predicts that the slope of $D(\omega)$ should be 2 in the intermediate regime $\omega_0 < \omega \ll \omega_*$, indicated by the triangle in Figure \ref{fig:Domegadata}a. We find reasonable agreement with this prediction, but larger packings are needed to test this definitively. Recent simulations \cite{Mizuno:2013} of a related model found $D(\omega) \sim \omega^{1.5}$ when $\omega_{BP} \approx 0$, which is also compatible with our data, and with earlier theory \cite{Grigera:2003}. 

To test our prediction for $\mu(e/e_c)$, equation \eqref{mu}, we repeat the numerical experiment above, but in 2D. We use 1000 bidisperse packings of 25600 disks, constructed with FIRE, at strains from $e_0=10^{-2}$ to $10^{-4}$. Rescaling contact forces by a factor $1-x$, we again explore a range of $e(x)/e_0$ from 0 to 1, and again we find $e_0/e_c=0.96$. The measured shear modulus $\mu$ is shown in Figure \ref{fig:mu}. In agreement with theory, the data collapse when rescaled by $\dz$, and drop by a factor of 2 between $e=0$ and $e=e_c$. 


These results are expected to persist at finite but low temperature. A nice example are colloidal glasses, where particles are hard spheres. In the glass phase, contacts can be defined   by considering those particles who collide with each other on a time scale much smaller than the relaxation time $\tau_\alpha$ where the system is liquid, but much larger than the typical collision time scale between two neighbors \cite{BritoWyart:2006,BritoWyart:2009}.  The contact strain is simply the mean distance between two particles. In such glasses one indeed finds marginal stability with  a boson peak frequency $\omega_{BP}\ll \omega_*$  \cite{BritoWyart:2009}.

One aspect of packing that we did not seek to capture is that the bulk modulus remains finite as $\dz \to 0$.  This property is not a generic feature of weakly-coordinated materials, as generically in random elastic networks the bulk and shear modulus scale identically. However, this point is well understood \cite{Wyart053}, and is due to the fact that the geometry of packings is such that contact forces must all be positive. This will thus be true when the potential is strictly repulsive.  To construct an effective medium  theory capturing this fact one could enforce that the bulk modulus in Eq.(\ref{G2}) is constant. However, we expect that this modification will mostly affect the speed of sound of compressive waves below $\omega_*$, and we leave this point for further investigation.

{\it  Silica:} Silica is the most common glass,  with  a very large boson peak. It has been argued \cite{matsi,Wyart053} that this is the case because silica is marginally connected \cite{kostya}.  Indeed in this glass (or more generally aluminosilicates) the forces within the tetrahedra SiO$_4$  are much stronger than the forces that act between them \cite{hammonds}: it is easier to rotate two linked tetrahedra than to distort one tetrahedron\footnote{For example the bending energy of Si-O-Si is roughly 10 times smaller than the stretching of the contact Si-O \cite{ff}.}. This suggests a model of such glass as an assembly of linked tetrahedra loosely connected at corners: this is the ``rigid unit modes'' (RUM) model \cite{heine}.  
Such  a tetrahedral network with completely flexible joints is marginally connected \cite{kostya}:  on the one hand each tetrahedron has 6 degrees of freedom (3 rotations and 3 translations). On the other hand, the 4 corners of a tetrahedron each bring 3 constraints shared by 2 tetrahedra, leading to 6 constraints per tetrahedron and thus $\delta z=0$. Within  our approach, the RUM model corresponds to $\delta z=0, e=0$ and must thus have  a flat density of states, and an infinite boson peak amplitude, as is indeed observed numerically \cite{dove}.


These predictions do not  describe well the spectrum of silica at low frequencies, where the weak interactions, in particular the bending of the Si-O-Si bond and the  Van der Waals interactions cannot be neglected.
These interactions imply that $e<0$.  For this case, we predict that the spectrum is characterized by one frequency scale only, as $\omega_0\approx\omega_*\approx\omega_{IR}\approx \omega_{BP}\propto \sqrt{-e}$, see Fig. \ref{fig:phasediagram} and Eq. \eqref{ee}. Using the stiffness of the Si-O-Si bending interaction obtained {\it ab initio} \cite{ff}, and the molecular mass to form a frequency, one obtains a crude estimate  $\omega_*\approx 1.4 THz$.  Our predictions are in agreement with measurements of the density of states in silica, which indeed present a plateau above the boson peak frequency at about $1THz$, see e.g. \cite{kostya}. Since the bending stiffness of Si-O-Si is roughly 200 times smaller than the stretching stiffness of the bond Si-O \cite{ff}, we estimate $\omega_* \approx 1/\sqrt{200} \sim 0.07$ in our units, indicating $e \approx -(0.07/c_2)^2 \sim -0.01$. Concerning transport, we predict that the Ioffe-Regel frequency and the boson peak are nearly identical, in agreement with experiments \cite{Baldi:2010}. We predict that above this frequency, the mode diffusivity displays a plateau.  This prediction enables to capture quantitatively the high temperature behaviour of thermal conductivity of silica using $e=-0.007$, as shown in Fig. \ref{fig:kappa}. 

These arguments apply equally to Germanium oxide. However, in amorphous Germanium or Silicon, a tetrahedral structure is also formed, but the joint between tetrahedra are not flexible at all. 
$\delta z$ is thus large in these systems, and our analysis thus predicts that the boson peak should be small. 

\begin{figure}[ht!] 
\begin{tikzpicture}[scale=1]
\clip (0,0) rectangle (8,6.5);
\node[above right]{\includegraphics[viewport=16 7 457 360,width=0.49\textwidth]{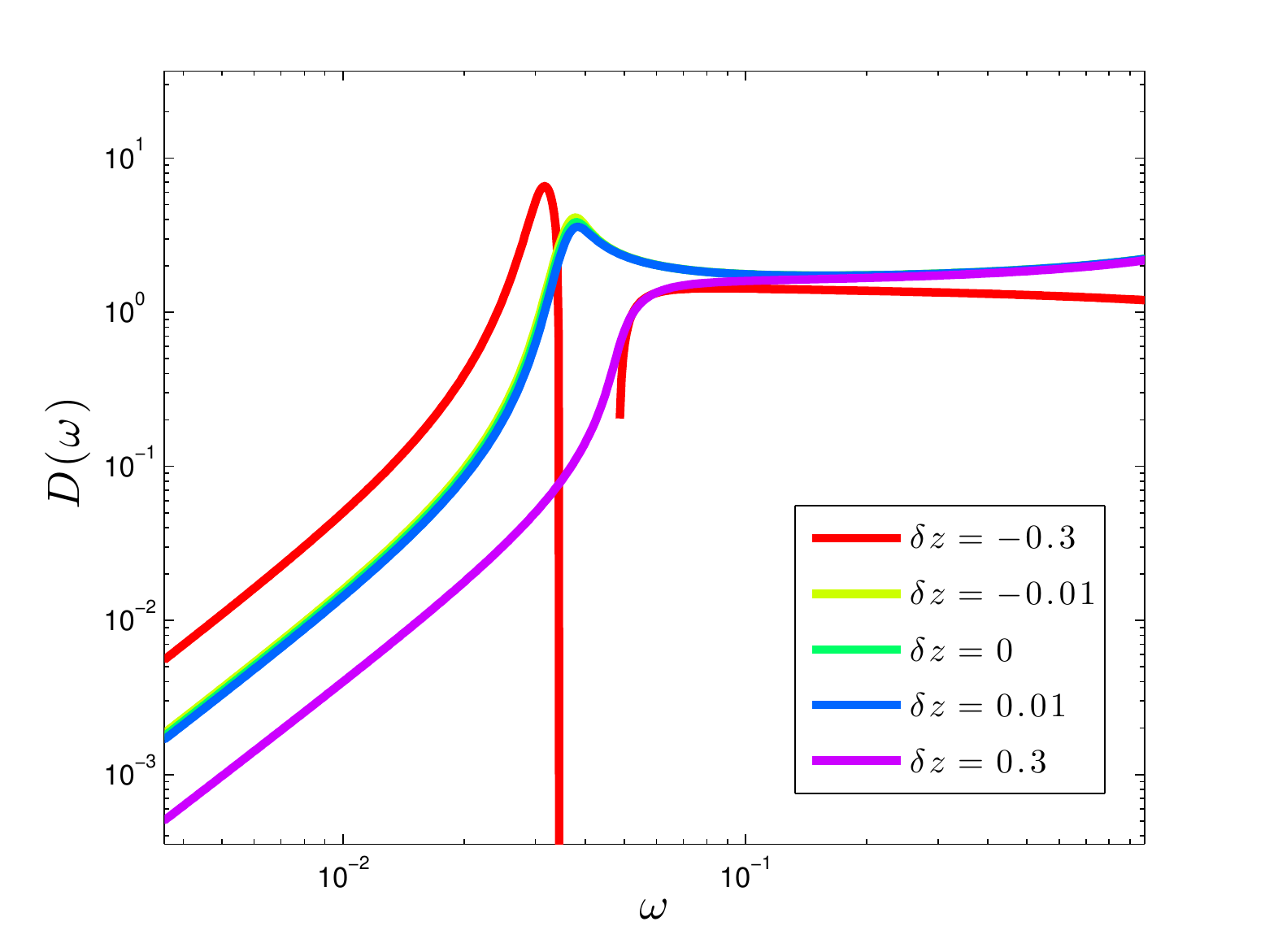}};
\end{tikzpicture}
\caption{\label{fig:chalc1} Density of states $D(\omega)$ for $e=-0.01$ at indicated values of $\dz$. }
\end{figure}
\begin{figure}[ht!] 
\begin{tikzpicture}[scale=1]
\clip (0,0) rectangle (8,6.5);
\node[above right]{\includegraphics[viewport=16 7 457 360,width=0.49\textwidth]{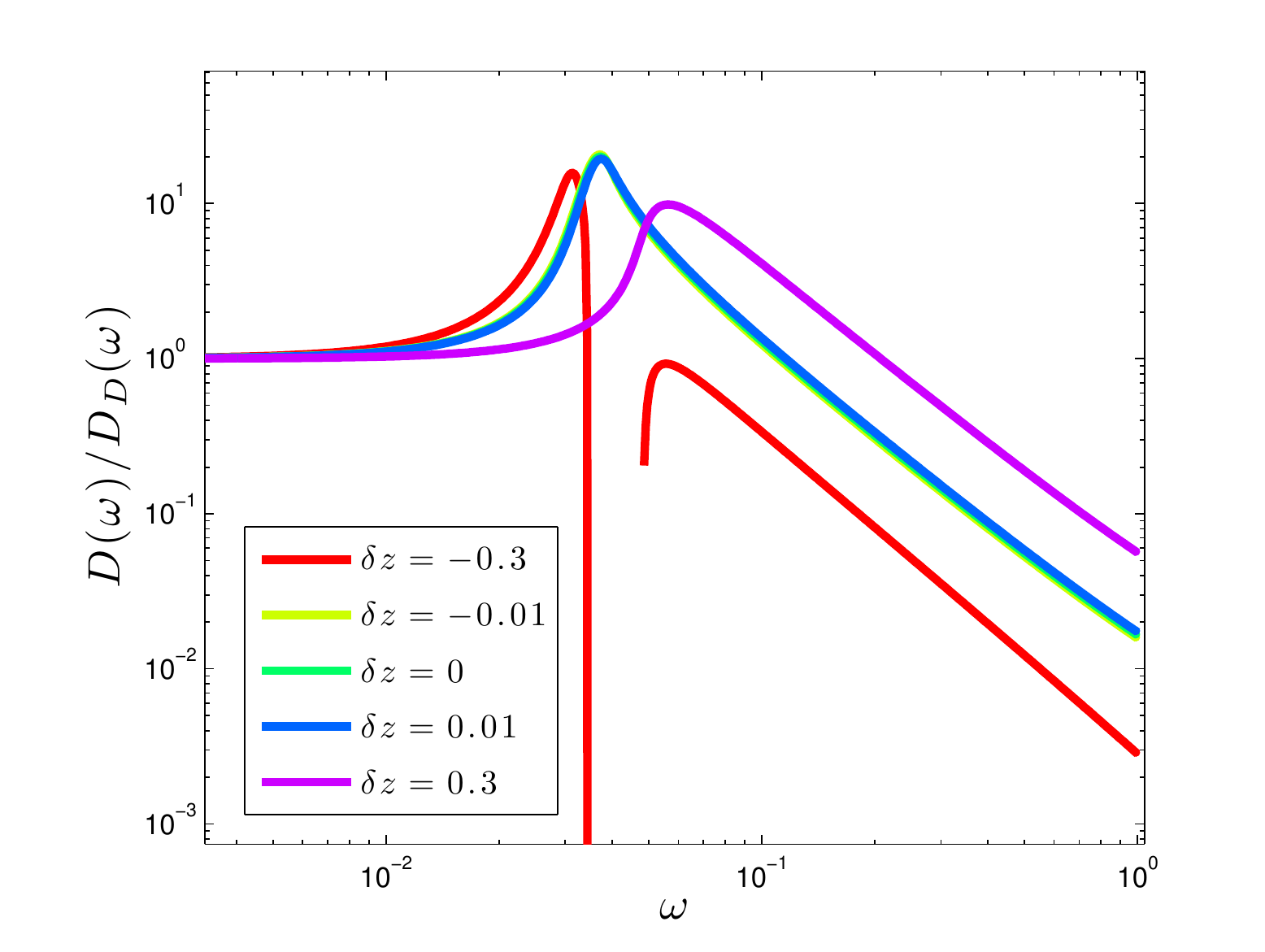}};
\end{tikzpicture}
\caption{\label{fig:chalc2} Reduced density of states $D(\omega)/D_D(\omega)$ for $e=-0.01$ at indicated values of $\dz$, where $D_D \sim \omega^2$ is the Debye density of states. }
\end{figure}

{\it Covalent Networks:} Network glasses are very convenient to test our predictions. In chalcogenides, for example, the connectedness can be changed continuously by considering compounds of elements of different valence, such as Se$_x$As$_y$Ge$_{1-x-y}$, whose valence is $z=2x+3y+4(1-x-y)$. In these systems, Phillips showed that $z_c=2.4$. When $\delta z<0$, Van der Waals interactions stabilize these materials, and their relative amplitude can be estimated from measurements of the dependence of the shear modulus with coordination \cite{Yan:2013}, from which one gets the order of magnitude $e\approx -0.01$. 

The normalized boson peak amplitude, $A_{BP}$, is the maximum of the reduced density of states, $D(\omega)/D_D(\omega)$, where $D_D(\omega)$ is the Debye density of states. The reduced density of states is plotted in Fig. \ref{fig:chalc2} for $e=-0.01$, showing a non-monotonic dependence of $A_{BP}$ on $\dz$: it has a maximum at $\dz=0$. This is consistent with simulations on elastic networks \cite{Yan:2013}.

For well-coordinated glasses with $\dz>0$ (lower right part of our phase diagram), our predictions are as as follows. The spectrum is characterized by one frequency scale only, very much like for silica. This frequency scale increases with coordination, such that very well-coordinated glasses have a small boson peak located at high frequency-- e.g. amorphous silicon, see discussion above. Above this frequency, the mode diffusivity is flat.

For under-coordinated  glasses with $\dz<0$, if $e=0$ effective medium predicts that the density of states present zero modes \cite{thorpe2,During:2013} and a gap up to a frequency $\omega_*\sim -\delta z$ \cite{During:2013}. In this regime, our EMT equations \eqref{EMTeq} apply, and can be solved numerically; the resulting $D(\omega)$ is shown in Fig. \ref{fig:chalc1}, and in reduced form in Fig. \ref{fig:chalc2}. However, note that the asymptotic solution derived in Appendix B does not apply; the relevant asymptotic solution will be discussed elsewhere.

When $e<0$, floppy modes get a finite frequency of order $\sqrt{|e|}$, so that the gap is present at intermediate frequencies only for $\delta z<-\sqrt{|e|}$. In the phase diagram of Fig. \ref{fig:phasediagram2}, this occurs to the left of the red line. When $\delta z > -\sqrt{|e|}$, no gap is present. The modes above $\omega_*$ are predicted to have a flat diffusivity.

\section{Discussion and Conclusion}

Many approaches to understanding the boson peak describe how an elastic instability is reached as a parameter is changed. Some features of vibrational properties are expected to be universal near such an instability, independent of the realism of the model \cite{Grigera:2002,Parisi:2003}. For example, several predictions of Schirmacher, where the control parameter is the amplitude of disorder \cite{Schirmacher:2006,Schirmacher:2007}, are similar to ours, including the transition of sound dispersion $\Gamma(\omega)$ from $\omega^4$ to $\omega^2$, and a shear modulus that drops by a factor of 2 as the instability is approached \cite{Marruzzo:2013}, in broad agreement with experiments and simulations. 

However, universality implies that such successes 
do not guarantee that the key structural aspects controlling the boson peak have been correctly identified. Several approaches propose to classify the vibrational properties of glasses by their amount of structural disorder. As discussed above, this classification cannot capture the similarity in the boson peak, in some glasses, to the boson peak in their crystalline counterparts. In that regard, observing that a dip in the speed of sound occurs at  the boson peak  frequency, as was done numerically in a Lennard-Jones glass \cite{Marruzzo13}, is not a strong support that fluctuations in the shear modulus are responsible for the peak. Predicting how the latter evolves as a parameter (such as density) may be more appropriate to distinguish theories, as was done in a Lennard-Jones in \cite{Xu:2007}.  Likewise, observed correlations between the boson peak  amplitude and the presence of large scale elastic heterogeneites do not imply  that vibrational anomalies are caused by large fluctuations in the structure. 
In a companion paper we show  that the length scale $l_c$ beyond which continuum elasticity breaks down, and at which fluctuations in elastic response are large, follows $l_c\sim 1/\sqrt{\omega_0}$ \cite{Lerner:2013b}.  A fundamental point is that in our model, this length scale {\it does not} enter in the static structure of the glass, but only in its response.  

We argue that with two parameters, the connectedness $z$ and the parameter $e$ that includes compression and the relative strength of weak interactions, specific non-trivial predictions can be made, for example on the relationship between different characteristic frequency scales. This approach captures both the qualitative features of sound dispersion near the boson peak, as well as the fact that the latter is similar in some glasses and in their corresponding crystals. We hope that our phase diagram, aimed at characterizing  emulsions, colloidal glasses (where several experimental measurements of the vibrational spectrum and microscopic structure support our views \cite{chen2010,Ghosh:2010,brujic2}), and covalent glasses will be a convenient starting point to classify a broader class of amorphous solids. Our predictions on network glasses could be tested experimentally by changing the microscopic structure in a systematic way (e.g. by monitoring the valence in chalcogenides and measuring transport properties).

One essential result concerns transport. We predict  that the mode diffusivity becomes frequency-independent independent of the glass, above the frequency scale $\omega_0$ where  the  density of states departs from Debye behavior, in agreement with numerical observations in sphere packings \cite{Vitelli:2010,Xu:2009}. For silica or well-coordinated covalent glasses (for which $\omega_0 \approx \omega_*$), this fact results from a cancellation:  above $\omega_*$ the modes' characteristic velocity increases as $\nu(\omega)\sim \sqrt{\omega}$ whereas their length scale decreases as $l_s(\omega)\sim 1/\sqrt{\omega}$, such that $d(\omega)\sim \ell(\omega) \nu(\omega)\sim \omega^0$. This prediction could be tested empirically by extracting the frequency-dependence of $l_s(\omega)$ and $\nu(\omega)$ from scattering data, as discussed above. 

Concerning the evolution of the shear modulus $\mu$ near an elastic instability, we find that the elastic moduli do not vanish, but only drop by a factor of 2 at instability. 
Our results are supported by packings of spheres at low pressure, which are close to an elastic instability, but where $\mu$ only mildly depends on pre-stress, as discussed above. Note that the factor 2 that bounds the evolution of $\mu$ only holds at fixed connectedness $z$, and can be larger if system is allowed to change coordination as well.   It remains to be seen if the proximity of an elastic instability  is the main cause for the evolution of $\mu$ with temperature in fragile liquids.  

\appendix 
\section{EMT} 

\renewcommand{\theequation}{A.\arabic{equation}}
\setcounter{equation}{0}

The quadratic energy expansion $\delta E = \sum_\alpha \delta E_\alpha = \langle \delta \bm{R} | \curlyM | \delta \bm{R} \rangle$ defines the dynamical matrix 
\def\nn{\nb_{\alpha} \! \otimes \! \nb_{\alpha}}
\eq{ \label{curlyM}
\curlyM = \!\sum_{\alpha} \left[ k_{\alpha} \nn - \frac{f_{\alpha}}{\sigma_\alpha} (\delb - \nn) \right] \curlyP_{\alpha},
}
where $\bm{n}_\alpha$ is a unit vector along the bond $\alpha$, $\delb$ is the unit tensor, and $f_{\alpha}$ is the force in the contact $\alpha$. Here $\curlyP_{\alpha} = \half (|i\rangle - |j\rangle)(\langle i | - \langle j |)$ is a projection operator for the contact $\alpha$ \cite{Kirkpatrick:1973}. The Green's function is $\Gb(\omega) = \big(\curlyM - m\omega^2\big)^{-1}$. 

To obtain the disorder-averaged Green's function $\overline{\Gb}$, we first write $k_\alpha = \kpa + (k_\alpha-\kpa)$, where $\kpa$ is an effective spring constant, and likewise $f_\alpha/\sigma_\alpha = e \kpe + e(k_{\alpha} - \kpe)$. The effective and fluctuating contributions in $\curlyM=\overline{\curlyM}+\delta \curlyM$ are collected into $\overline{\curlyM}$, and $\delta \curlyM$, respectively. $\overline{\curlyM}$ is constructed from $\curlyM$ by making replacements $k_\alpha \to \kpa$ and $f_\alpha/\sigma_\alpha \to \kpe e$, and similarly for $\delta \curlyM$. The Green's function can then
%
%
%
be written as $\Gb = \overline{\Gb} + \overline{\Gb} \curlyT \overline{\Gb}$, where $\overline{\Gb}=(\overline{\curlyM}-m\omega^2)^{-1}$ is the effective Green's function, and $\curlyT$ is known as the transfer matrix. The transfer matrix is written as $\curlyT = -\delta \curlyM (1 + \Gbo \delta\curlyM)^{-1} = -\delta \curlyM \sum_{n \geq 0} (-\Gbo \delta \curlyM)^n$. Since $\curlyP$ is a local operator, $\curlyT$ can efficiently be organized by its contributions from increasing numbers of contacts, viz., 
\eq{
\curlyT = \sum_\alpha \curlyT_\alpha + \sum_\alpha \sum_{\beta \neq \alpha} \curlyT_{\alpha} \overline{\Gb} \curlyT_{\beta} + \ldots 
}
We find that
 \eq{
 \curlyT_\alpha & = \curlyP_\alpha \left[  \frac{\kpa-k_\alpha}{1-(\kpa-k_\alpha) G^\parallel} \nn \right. \notag  \\
 & \qquad \left. - \frac{e \kpe-e k_\alpha}{1+(e \kpe-e k_\alpha) G^\perp} \big(\delb  - \nn \big) \right],
 }
where $G^\parallel$ and $G^\perp$ are the longitudinal and transverse components of the Green's function on a bond, which can be written in terms of $\langle \alpha | \equiv \langle i | - \langle j|$ as
\eq{
G^\parallel & = 2 \nb_\alpha \cdot \langle \alpha | \overline{\Gb} | \alpha \rangle \cdot \nb_\alpha \notag \\
G^\perp & = 2 \tr( \langle \alpha | \overline{\Gb} | \alpha \rangle) - G^\parallel,
}
where $\tr$ is trace. By isotropy and homogeneity of the effective lattice, $G^\parallel$ and $G^\perp$ are independent of $\alpha$. To obtain $\overline{\Gb}$, we should choose effective constants such that $\overline{\curlyT} = 0$. 
%
 In EMT this is approximated by $\overline{\curlyT_\alpha} = 0$.\footnote{The leading error is a correlation of the form $\overline{\curlyT_\alpha \otimes \curlyT_\alpha}^2 \overline{ \langle \alpha | \Gb | \beta \rangle }^3$ 
 \cite{Kirkpatrick:1973}.} This requires
\eq{
0 = \overline{ \frac{\kpa-k_\alpha}{1-(\kpa-k_\alpha) G^\parallel} } = \overline{ \frac{e \kpe-e k_\alpha}{1+(e \kpe-e k_\alpha) G^\perp} }.
}
As discussed in the main text, random dilution of the lattice is modelled by setting $k_\alpha = k_0$ with probability $P=z/z_0$ and $k_\alpha=0$ with probability $1-z/z_0$. This leads to EMT equations 
\eq{ 
G^\parallel & = \frac{\kpa-P}{\kpa(\kpa-1)}, \label{emt1} \\
G^\perp & = -\frac{\kpe-P}{e \kpe(\kpe-1)}, \label{emt2}
}
where we have taken units such that the bare spring constant $k_0=1$. Using the identity $\delb = \langle i | \Gbo (\overline{\curlyM} - m \omega^2) | i \rangle$ and homogeneity and isotropy of the lattice \cite{Kirkpatrick:1973, Wyart:2010}, one can derive an exact identity
\eq{
\frac{z_0}{2d} \big( G^\parallel \kpa - e G^\perp \kpe \big) = \left( 1 + \frac{m \omega^2}{d} \tr(\overline{\Gb}(0,\omega)) \right).
}
For simplicity, we will neglect the difference between the longitudinal and transverse speed of sound, the consequences of which are discussed in the conclusion. Then, restoring isotropy by averaging \eqref{Gb} over orientation of the lattice with respect to the laboratory frame, this assumption implies
\eq{ \label{G1}
G^\parallel = G^\perp = \frac{2d}{z_0} \frac{1}{\kpa-e \kpe} \left( 1 + \frac{m \omega^2}{d} \tr(\overline{\Gb}(0,\omega)) \right),
}
where $\Gbo$ is the disorder-averaged Green's function. 

\section{Asymptotic solution} 
\renewcommand{\theequation}{B.\arabic{equation}}
\setcounter{equation}{0}

Here we derive the asymptotic solution for $\kpa$ and $\kpe$ when $\dz \ll 1$ and $\omega \lesssim \dz$. We expect that there is a critical strain $e_c(\delta z)$ such that solutions fail to exist for $e > e_c(\delta z)$, so it is natural to look for solutions in the variables $\delta z$ and $e' = e/e_c(\delta z)$. 

When $e=0$, previous work \cite{Wyart:2010} shows that $\kpa \sim \dz$, so we look for a solution
\eq{ \label{expansion}
\kpa & = \kpa_0 \delta z + \kpa_1 (\delta z)^\gamma + \ldots \\
k^\perp & = \kpe_0 (\delta z)^\alpha + \kpe_1 (\delta z)^\beta + \ldots \\
e_c & = e_1 (\dz)^\eta + e_2 (\dz)^\zeta + \ldots
}
We also rescale $\omega= \dz \; \omega'$. When $\delta z \ll 1$, the transverse stiffness $e \kpe$ should be much smaller than the normal stiffness, so $\alpha + \eta > 1$. In fact, simulations indicate that $\alpha + \eta= 2$. We assume this in what follows, and derive it at the end of this section.

In $d=3$ the EMT equations are
\eq{ \label{EMTeq}
G^\parallel & = \frac{\kpa-z/z_0}{\kpa(\kpa-1)}, \qquad G^\perp  = -\frac{\kpe-z/z_0}{e \kpe(\kpe-1)}, \notag \\
G^\parallel & = G^\perp = \frac{6}{z_0 \Delta k} \left[1 + \frac{A_1 \omega^2}{\Delta k} \right. \notag \\
& \left. - \frac{A_1 \omega^3}{\Lambda (\Delta k)^{3/2}} \mbox{atanh}(\Lambda\sqrt{\Delta k}/\omega) \right], 
}
where $\Delta k = \kpa - e \kpe$, $\Lambda$ is a Debye cutoff, and 
$A_1 = \Lambda z_0/(6\pi^2)$. 
With the above scalings, $\sqrt{\Delta k}/\omega \sim 1/\sqrt{\dz}$ so atanh can be expanded around infinity, giving
\eq{
\frac{z_0}{6} & \Delta k G^\parallel -1 = \frac{A_1 \omega^2}{\Delta k} + \frac{i A_2 \omega^3}{(\Delta k)^{3/2}} + \OO(\dz^2) \notag \\
& = \dz A_1 \frac{(\omega')^2}{\kpa_0} - \dz^{\gamma} A_1 \frac{(\omega')^2 \kpa_1}{(\kpa_0)^2} + i A_2 \dz^{3/2} \frac{(\omega')^3}{(\kpa_0)^{3/2}} + \ldots
}
with $A_2=\pi A_1/(2 \Lambda)$. This must be equated to
\eq{
\frac{z_0}{6} \Delta k G^\parallel -1 & =\dz \left( -a \kpa_0 - \frac{e_1 e' \kpe_0}{\kpa_0} + \frac{1}{2d} \right) \notag \\
& \qquad + \dz^\gamma \kpa_1 \left( -a + \frac{e_1 e' \kpe_0}{(\kpa_0)^2} \right) + \ldots,  
}
with $a=z_0/(2d)-1$, assuming $\beta -\alpha > \gamma-1$ and $\zeta-\eta > \gamma -1$, verified below. The $O(\dz)$ equation gives
\eq{ \label{kpasol}
\kpa_0 = C_1 \omega'_* \pm C_1 \sqrt{(\omega'_0)^2 - (\omega')^2},
}
with $C_1= \sqrt{A_1/a}$, $\omega'_* = 1/(4da C_1)$, and $\omega'_0(e') = \omega_*' \sqrt{1 - 16 a d^2 e_1 e' \kpe_0}$. The next order must be $\gamma=3/2$, giving
\eq{
\kpa_1 = \frac{\mp i C_2 (\omega')^3}{\sqrt{(\omega'_0)^2 - (\omega')^2} \sqrt{\omega_*' \pm \sqrt{(\omega'_0)^2 - (\omega')^2}}},
}
with $C_2 = A_2/(2aC_1^{3/2})$. The transverse term is
\eq{ \label{trans}
\frac{z_0}{6} \Delta k G^\perp = \frac{z_0}{6} \left( 1 - \frac{\kpa}{e \kpe} \right) \frac{\kpe-z/z_0}{\kpe-1}
}
To match with the above, the RHS must be $1 + \OO(\dz)$. This implies $\alpha=0$, $\eta=2$, and $\kpe_0 = 2d/z_0$. We find
\eq{
\frac{z_0}{6} \Delta k G^\perp = \frac{z_0}{6} \left( \dz - \frac{\kpa_0}{e_1 e'  \kpe_0} \right) \frac{2d/z_0 - \dz^{\beta-1} \kpe_1}{2da/z_0} + \ldots, \notag
}
implying $\beta=1$ and $\kpe_1 = 1/z_0 + ae_1 \kpa_0/(1+a)^3$. Hence we have
\eq{
\Delta k = \dz \kpa_0 + \dz^{3/2} \kpa_1 + O(\dz^2)
}
The density of states is
\eq{
D(\omega) & =(2\omega/\pi) \mbox{ Im[tr[}\overline{\Gb}(0,\omega)]] \notag \\
& = \frac{z_0}{\pi \omega} \mbox{ Im[}\Delta k G^\parallel]
}
In order to have a non-negative density of states $D(\omega)$ as $\omega \to 0$, we must have Im$[\kpa_1]<0$ as $\omega \to 0$, indicating that we must take the positive root in \eqref{kpasol}. Finally, to determine $e_1$, we must use the fact that $e_c(\dz)$ corresponds to the critical pressure. Instability is signalled by movement of an eigenvalue $\lambda=\omega^2$ to negative values, hence as $\omega \to 0$, $D(\lambda)=D(\omega)/(2\omega) \to 0$ at the critical pressure $e'=1$, and $D(\lambda)>0$ when $e'>1$. This leads to Im$[1/\kpa_0(0)]=0$, to order $\OO(\dz)$. This implies $\omega_0'(e'=1)=0$, or $e_1 =  z_0/(4a(2d)^3)$. It can be seen that the $\OO(\dz^{3/2})$ term does not require a corresponding term in $e_c$, verifying that $\zeta > \eta + \gamma -1  = 5/2$, assumed above. Rewriting frequencies in unscaled variables, $\omega = \dz \omega'$, and keeping only the leading terms for $D(\omega)$, these expressions then reproduce what is given in the main text in equation \eqref{Domega}, with $c_1= (4daC_1)^{-1}$ and $c_2=c_1/\sqrt{e_1}$. It is notable that these first terms in an asymptotic solution, which reproduces all of the scaling behaviour discussed in the main text, only used the $\omega=0$ and singular parts of the Green's function \eqref{G2}; this partially justifies the simple continuum expression used. 

In this derivation, we assumed $\alpha + \eta=2$. To see why this must be true, consider \eqref{expansion} but with 
\subsection{Case 1: $\alpha+\eta=1$}
From the expansion of atanh, we will again have $(z_0/6) \Delta k G^\parallel -1 \sim \dz$. But 
\eq{
(z_0/6) \Delta k G^\parallel -1 & = -e_1 e' \kpe_0/\kpa_0 + O(\dz) \notag \\
& \qquad  + O(\dz^{\gamma-1}) + O(\dz^{\beta-\alpha}), 
}
so that equating these will lead to $e_1 \kpe_0=0$. Hence $\alpha+\eta=1$ is the wrong scaling. The same argument also excludes $\alpha+\eta < 1$. 
\subsection{Case 2: $\alpha+\eta > 1$} 

For a general $\alpha+\eta>1$, we will have $(z_0/6)\Delta k G^\parallel -1 \sim \dz$, and equation \eqref{trans} still holds, implying that $\alpha=0$ and $\kpe_0 = 2d/z_0$. Then
\eq{
\Delta k G^\perp \sim \left( \dz - \dz^{2-\eta} \frac{\kpa_0}{e_1 e'  \kpe_0} \right) \big( 2d/z_0 - \dz^{\beta-1} \kpe_1 \big). \notag
}
Since $\eta>1$, the leading terms are $\OO(\dz^{2-\eta})$ and $\OO(\dz^{1+\beta-\eta})$, which must be $\OO(1)$ to match with the other equations. If $\eta=2$, we're done, so consider $\beta=\eta-1$. The leading terms are then $\OO(\dz^{2-\eta})$ and $\OO(\dz^{0})$. If these are not equal, the system is overdetermined, so $\eta=2$. 

\section{Diffusivity} 
\renewcommand{\theequation}{C.\arabic{equation}}
\setcounter{equation}{0}

Energy diffusivity $d(\omega)$ can be calculated with the Kubo-Greenwood formula for the thermal conductivity \cite{Vitelli:2010,Xu:2009}. For a finite system, this leads to
\eq{ \label{dkubo}
d(\omega) = \frac{\pi}{12 m^2 \omega^2} \sum_{\omega' \neq \omega} \frac{(\omega+\omega')^2}{4\omega\omega'} |\bm{\Sigma}_{\omega \omega'}|^2 \tilde{\delta}(\omega-\omega'),  
}
where the sum is over eigenvalues ${\omega'}^2$ of $\curlyM$. Here the vector heat-flux elements are
\eq{
\bm{\Sigma}_{\omega \omega'} = \sum_{i,j} (\rb_i-\rb_j) \bm{\psi}_\omega^i \cdot \curlyM_{ij} \cdot \bm{\psi}_{\omega'}^j,
}
with $\rb_i$ the center of particle $i$, $\bm{\psi}_\omega^i$ the (vector) eigenvector of $\curlyM$ associated to $\omega$, and $\tilde{\delta}$ is a smoothed $\delta$-function, whose width should be taken to zero at the end of the calculation \cite{Vitelli:2010,Xu:2009}. We write the expression for $\curlyM$, equation \eqref{curlyM}, as $\curlyM = \sum_\alpha \bm{m}_\alpha \curlyP_\alpha$, where $\bm{m}_\alpha$ is the contribution from contact $\alpha$, a symmetric $3\times 3$ matrix. Then
\newcommand{\kron}{\!\otimes\!}
\eq{
\bm{\Sigma}_{\omega \omega'} = \half \sum_\alpha (\rb_i-\rb_j) \left[ \bm{\psi}_\omega^i \kron \bm{\psi}_{\omega'}^j - \bm{\psi}_\omega^j \kron \bm{\psi}_{\omega'}^i \right] : \bm{m}_\alpha,
} 
where `$:$' indicates two tensor contractions, and $\alpha = \langle i j \rangle$. It is clear from this expression that $\bm{\Sigma}_{\omega \omega'}$ is zero when $\omega'=\omega$, but the $\tilde{\delta}(\omega-\omega')$ factor in \eqref{dkubo} implies that modes with any finite frequency difference do not contribute to $d(\omega)$. Hence, in the thermodynamic limit, only modes which are infinitesimally close in frequency can contribute to $d(\omega)$. Making the replacement $\rb_i-\rb_j=\nb_\alpha$, the squared magnitude of $\bm{\Sigma}_{\omega \omega'}$ is 
\eq{
|\bm{\Sigma}_{\omega \omega'}|^2 & = \ffrac{1}{4} \sum_{\alpha,\beta} \nb_\alpha \!\cdot\! \nb_\beta \left[ \bm{\psi}_\omega^i \kron \bm{\psi}_{\omega'}^j - \bm{\psi}_\omega^j \kron \bm{\psi}_{\omega'}^i \right] : \bm{m_}\alpha \notag \\
& \qquad \times \left[ \bm{\psi}_\omega^k \kron \bm{\psi}_{\omega'}^\ell - \bm{\psi}_\omega^\ell \kron \bm{\psi}_{\omega'}^k \right]^\dagger : \bm{m}_\beta^\dagger,
}
where $\beta = \langle k \ell \rangle$, and $\dagger$ denotes complex conjugate. We are interested in the disorder average of this quantity. It was previously established in numerical simulations that modes of close but unequal frequency are uncorrelated \cite{Vitelli:2010}. We can then obtain an EMT estimate of $\overline{|\bm{\Sigma}_{\omega \omega'}|^2}$ using (i) the identity $\langle \bm{\psi}_{\omega}^j \kron {\bm{\psi}_{\omega}^k}^\dagger \rangle = -2\mbox{Im}[\Gb(\rb_j-\rb_k,\omega)] \omega/(3\pi N D(\omega))$, and (ii) replacing $\Gb$ and $\bm{m}_\alpha$ by their EMT values $\Gbo$ and $\overline{\bm{m}}_\alpha$. Using the fact that $\Gbo \propto \delb$, we find
\eq{
& \overline{|\bm{\Sigma}_{\omega \omega'}|^2} \approx \frac{1}{\pi^2 N^2}\frac{\omega}{D(\omega)} \frac{\omega'}{D(\omega')} \sum_{\alpha,\beta} \nb_\alpha \!\cdot\! \nb_\beta \; \tr( \overline{\bm{m}_\alpha} \cdot \overline{\bm{m}_\beta}^\dagger ) \notag \\
&\quad  \times \left[ I^\omega_{ik} I^{\omega'}_{j\ell} - I^\omega_{i\ell} I^{\omega'}_{jk} - I^\omega_{jk} I^{\omega'}_{i\ell} + I^\omega_{j\ell} I^{\omega'}_{ik} \right],
}
where $I^\omega_{ik} = \tr(\mbox{Im}[\Gbo(\rb_i-\rb_k,\omega)])$. It is now possible to let $\omega \to \omega'$. Then $\sum_{\omega'} \tilde{\delta}(\omega-\omega') \to 3N D(\omega)$. We find
\eq{ \label{dest}
d(\omega) \approx \!\frac{(2\pi)^{-1}}{N D(\omega)} \sum_{\alpha,\beta} \nb_\alpha \!\cdot\! \nb_\beta \; \tr( \overline{\bm{m}_\alpha} \cdot \overline{\bm{m}_\beta}^\dagger ) \left[ I^\omega_{ik} I^{\omega}_{j\ell} - I^\omega_{i\ell} I^{\omega}_{jk} \right].
}
This has both `diagonal' $\alpha=\beta$ and `off-diagonal' $\alpha \neq \beta$ contributions, denoted $d_{d}(\omega)$ and $d_{od}(\omega)$, respectively. Using $\tr( \overline{\bm{m}_\alpha} \cdot \overline{\bm{m}_\alpha}^\dagger ) = |\kpa|^2 + 2 e^2 |\kpe|^2 \approx |\Delta k|^2$, we find the diagonal contribution to be
\eq{
d_d(\omega) \approx \frac{|\Delta k(\omega)|^2}{2\pi N D(\omega)} \sum_{\alpha} \left[ I^\omega_{ii} I^{\omega}_{jj} - I^\omega_{ij} I^{\omega}_{ji} \right]
}
The quantity in parentheses is
\eq{
 I^\omega_{ii} I^{\omega}_{jj} - I^\omega_{ij} I^{\omega}_{ji} & = (I^\omega_{ii} - I^\omega_{ij})( I^\omega_{ii} + I^{\omega}_{ji}) \notag \\
 & = \frac{3^2}{4^2} \mbox{Im}[G^\parallel] \mbox{Im}\left[\frac{8}{3} \tr(\Gbo(0,\omega) - G^\parallel \right]
}
Using the asymptotic solution and keeping only leading terms, we find
\eq{
 I^\omega_{ii} I^{\omega}_{jj} - I^\omega_{ij} I^{\omega}_{ji} & \approx - \frac{3}{2} \frac{\pi}{2 \omega} \frac{\mbox{Im}[\Delta k] \mbox{Re}[G^\parallel]}{\mbox{Re}[\Delta k]} D(\omega) \notag \\
 & \approx - \frac{9\pi}{2\omega} \frac{\mbox{Im}[\Delta k]}{|\Delta k|^2} D(\omega)
}
and hence
\eq{
d_d(\omega) \approx -\frac{9 z}{8} \frac{\mbox{Im}[\Delta k]}{\omega}.
}
This can be written in terms of the macroscopic scales introduced in the main text. For brevity we omit writing the dependence on $\omega$. Using
\eq{
\mbox{Re}[\Delta k] & = \nu^2 \frac{n^2(n^2-1)}{(n^2+1)^2} \\
-\mbox{Im}[\Delta k] & = \nu^2 \frac{2n^3}{(n^2+1)^2},
}
we find
\eq{
d_d(\omega) \approx C_6 \ell_s \nu \frac{4n^2}{(n^2+1)^2},
}
with $C_6=9z/(16z_0)$. The off-diagonal contribution is more involved; here we look only for the dominant terms. Summing over contacts in \eqref{dest}, the only terms which survive are those which are symmetric both in $i$ and $j$, and in $k$ and $\ell$; i.e., the orientation of the contacts $\alpha$ and $\beta$ does not matter. By gradient expansion, the leading term involves the factor
\eq{
I^\omega_{ik} I^{\omega}_{j\ell} - I^\omega_{i\ell} I^{\omega}_{jk} = \nb_\alpha \cdot \nabla I^\omega_{ik} \nb_\beta \cdot \nabla I^\omega_{ik} + \ldots 
}
The Green's function needed for $I^\omega_{ik}$ is taken from its asymptotic large $r_{ik}$ behaviour, equation \eqref{Gdecay}. Since this depends only on $r$, we find, using $\tr( \overline{\bm{m}_\alpha} \cdot \overline{\bm{m}_\beta}^\dagger ) = |\Delta k|^2 (\nb_\alpha \cdot \nb_\beta)^2 + \OO(\dz^3)$,  
\eq{
d_{od}(\omega) \approx\! \frac{(2\pi)^{-1}|\Delta k|^2}{N D(\omega)} \sum_{\alpha,\beta} (\nb_\alpha \!\cdot\! \nb_\beta)^3 \nb_\alpha\! \cdot \bm{\hat{r}}_{ik} \; \nb_\beta\! \cdot \bm{\hat{r}}_{ik} \left[ \frac{\p I^\omega_{ik}}{\p r_{ik}} \right]^2
}
Assuming contact orientations are uncorrelated with $\bm{r}_{ik}$, this is
\eq{
d_{od}(\omega) & \approx \frac{(2\pi)^{-1}|\Delta k|^2}{V N D(\omega)} \left( \int_1^\infty dr \; r^2 \left[ \frac{\p I^\omega_{r}}{\p r} \right]^2 \right) \notag \\
& \qquad \times \left( \int_0^{4\pi} d\Omega \; \bm{\hat{r}} \kron \bm{\hat{r}} \right) : \sum_{\alpha,\beta} \nb_\alpha \kron \nb_\beta (\nb_\alpha \!\cdot\! \nb_\beta)^3,
}
where $V$ is the domain volume, and we integrate from $r=1$. Now we use
\eq{
\int_0^{4\pi} d\Omega \; \bm{\hat{r}} \kron \bm{\hat{r}} = \frac{4\pi}{3} \delb
}
and, for an isotropic material,
\eq{
\sum_{\alpha,\beta} (\nb_\alpha \!\cdot\! \nb_\beta)^4 = \left[\frac{Nz}{2}\right]^2 \frac{1}{\pi} \int_0^{\pi} \cos^4(\theta) d\theta = \frac{3 N^2 z^2 }{32}.
}
Using \eqref{Gdecay}, the nontrivial integral is 
\eq{
& \int_1^\infty dr \; r^2 \left[ \frac{\p I^\omega_{r}}{\p r} \right]^2  \notag \\
& \qquad = 9 C_5^2 \int_1^\infty dr \; r^2 \; \mbox{Im}^2 \left[ \frac{1}{r \Delta k} \left(g(\omega)-\frac{1}{r}\right) e^{g(\omega)r} \right],
} 
with $g(\omega)=i\omega/\nu(\omega) -1/\ell_s(\omega)$. Expanding Im$^2$, only one term does not have rapid oscillations. Keeping only leading terms, we finally find
\eq{
\int_1^\infty dr \; r^2 \left[ \frac{\p I^\omega_{r}}{\p r} \right]^2  & \approx \frac{9 C_5^2}{8 |\Delta k|^2} |g(\omega)|^2 \ell_s
}
and hence
\eq{
d_{od}(\omega) \approx \frac{9C_5^2 \rho z^2}{128} \frac{\omega^2 \ell_s}{\nu^2 D(\omega)} \left(1+\frac{1}{n^2}\right).
}
The density of states can be written
\eq{
D(\omega) = \frac{12A_1}{\pi^2} \frac{1}{\ell_s \nu} \left[ \pi + \frac{n^2-1}{\Lambda \ell_s}\right],
}
so that
\eq{
d_{od}(\omega) \approx C_7 \ell_s \nu \left(1+\frac{\pi \Lambda \ell_s-1}{n^2}\right)^{-1} \left(1+\frac{1}{n^2}\right),
}
with $C_7=3\pi^2\Lambda C_5^2 \rho z^2/(A_1 2^9)$. Assembling results into $d(\omega) \approx d_d(\omega) + d_{od}(\omega)$, we reproduce \eqref{domega} in the main text. We have assumed that (i) modes of unequal frequency are uncorrelated, (ii) the disorder average of a product (in particular, of $\Gb$ and $\bm{\hat{m}}^\alpha$) is equal to the product of their effective medium expressions, and (iii) we have kept only leading terms in $\dz$.

\begin{acknowledgments}

We thank Jie Lin, Le Yan and Marija Vucelja for discussions, and Tom Lubensky for sharing insights on effective medium with pre-stress. MW acknowledges support from NSF CBET Grant 1236378, NSF DMR Grant 1105387, and MRSEC Program of the NSF DMR-0820341 for partial funding. GD~acknowledges support from CONICYT PAI/Apoyo al Retorno 82130057.

\end{acknowledgments}
\bibliography{Wyartbib}

\end{document}